\documentclass[3p]{elsarticle}
\usepackage{graphicx}% Include figure files
\usepackage{dcolumn}% Align table columns on decimal point
\usepackage{bm}% bold math
\usepackage{lineno}
\usepackage{epsfig}
\usepackage{amsmath}
\usepackage{caption}
\usepackage{verbatim} %In order to be able to block comment using \begin{comment} etc.
\usepackage{subfigure}

\newcommand{\no}[1]{}

\newcommand\Rey{\mbox{\textit{Re}}}  % Reynolds number
\newcommand\Har{\mbox{\textit{Ha}}}  % Hartmann number
\makeatletter
\renewcommand\p@subfigure{\thefigure}
\makeatother

\journal{Journal of Computational Physics}

\begin{document}

\begin{frontmatter}

\title{ A hybrid finite difference/boundary element procedure for the simulation of turbulent MHD duct flow at finite magnetic Reynolds number }

%% Group authors per affiliation:
\author{Vinodh Bandaru}
%\author{Vinodh Bandaru\corref{cor}}
%\cortext[cor]{Corresponding author}
%\ead{vinodh-kumar.bandaru@tu-ilmenau.de}
\author{Thomas Boeck}
\author{Dmitry Krasnov}
\author{J\"{o}rg Schumacher}
\address{Institut f\"{u}r Thermo- und Fluiddynamik, Technische Universit\"{a}t Ilmenau, Postfach 100565, D-98684 Ilmenau, Germany}

\begin{abstract}
 A conservative coupled finite difference-boundary element computational procedure for the simulation of turbulent magnetohydrodynamic flow in a straight rectangular duct at finite 
magnetic Reynolds number is presented. The flow is assumed to be periodic in the streamwise direction and is driven by a mean pressure gradient. The duct walls are considered to be 
electrically insulating. The co-evolution of the velocity and magnetic fields as described respectively by the Navier-Stokes and the magnetic induction equations, together with the 
coupling of the magnetic field between the conducting domain and the non-conducting exterior is solved using the magnetic field formulation. The aim is to simulate localized magnetic fields
interacting with turbulent duct flow. Detailed verification of the implementation of the numerical scheme 
is conducted in the limiting case of low magnetic Reynolds number by comparing with the results obtained using a quasistatic approach that has no coupling with the exterior. 
The rigorous procedure with non-local magnetic boundary conditions is compared versus simplified pseudo-vacuum boundary conditions and the differences are quantified. Our first direct numerical
simulations of turbulent Hartmann duct flow at moderate magnetic Reynolds numbers and a low flow Reynolds number show significant differences in the duct flow turbulence, even at low 
interaction level between the flow and magnetic field.
\end{abstract}


\end{frontmatter}

\section{Introduction}
Turbulent conducting flows in the presence of a magnetic field occur in nature as well as in various industrial processes. In nature, it is observed in the generation of a magnetic field in the 
core of the Earth and in the Sun by the dynamo action and in the formation of other stars and galaxies \cite{Moffat-book}. Industrial processes include magnetic damping during the continuous 
casting of steel and aluminium, electromagnetic pumping of liquid metals, 
liquid metal blankets for future nuclear fusion power applications and material processing \cite{Smolentsev:2008,Davidson:1999}. These flows can be broadly distinguished on the 
basis of a non-dimensional parameter, the magnetic Reynolds number defined as
\begin{equation}
R_{m} = \frac{UL}{\lambda},
\end{equation}
where $U$ and $L$ are the charateristic velocity and length scales in the flow and $\lambda$ is the magnetic diffusivity of the fluid given by $\lambda = \left( \mu_{0}\sigma\right)^{-1}$,
$\mu_{0}$ and $\sigma$ being the magnetic permeability of free space and the electrical conductivity of the fluid respectively. $R_{m}$ is a measure of the relative magnitude of advection 
to the diffusion of magnetic field in the flow. Astrophysical MHD and the geodynamo fall into the category of advection dominated high $R_{m}$ flows ($R_{m}\gg1$) whereas most industrial
 flows involve moderate to low magnetic Reynolds numbers. The goal of this paper is to present a computational procedure for direct numerical simulations (DNS) of MHD duct flow at
 finite magnetic Reynolds number. By finite $R_{m}$, we imply the case $R_{m}\sim1$, when both magnetic advection and diffusion play an important role in determining the turbulent flow 
behaviour. In this regime, the coupling between the flow and the magnetic field is significant and the effect of their interactions on turbulence is one of the primary motivations for 
this work. Furthermore, an important application is in the technology of Lorentz Force Velocimetry (LFV), which involves reconstruction of velocity fields in hot and aggresive conducting 
flows (e.g., molten metals) by measuring the electromagnetic forces acting on magnet systems placed in the vicinity of the flow \cite{Thess:2006}. Transient MHD flows can also be associated
 with finite $R_{m}$ effects \footnote{In transient flows, $R_m$ becomes finite primarily due to the small inherent time scales rather than high flow velocities. The magnetic Reynolds
number defined as $R_{m} = L^2/\left(\tau_{adv}\lambda\right)$, where $\tau_{adv}$ is the advection time scale, becomes relevant in this context.} and hence proper understanding of fully 
resolved three-dimensional dynamics of turbulent flows in this regime is crucial in establishing an accurate and robust LFV.

Being frequently encountered in experimental studies of MHD and also in industrial applications, we choose the rectangular duct flow configuration for our study. Earlier studies of finite $R_{m}$ MHD turbulence have mostly been
performed in the periodic box setting (see for e.g. \cite{Oughton:1994,Knaepen:2004}). There are few existing studies of MHD turbulence at finite $R_{m}$ that include the presence 
of a mean shear with wall boundaries (e.g., \cite{Hamba:2010}). The main challenge in the numerical computation of finite $R_{m}$ MHD flows is the problem of magnetic boundary conditions that ensure proper 
matching of the magnetic field in the interior with that in the insulating exterior. This arises due to the fact that when $R_{m}$ is finite, the secondary magnetic field is non-negligible
and the equations governing it in the interior and exterior are different. In the case of spectral simulations in spherical geometries (as is the case with planets and stars), this 
problem is circumvented by poloidal-toroidal decomposition of the magnetic field and the use of expansions in spherical harmonics \cite{Christensen:2001}. Such a procedure leads to 
boundary conditions that are decoupled for each harmonic. Similar simplification of boundary 
conditions is possible in configurations with two periodic directions like that of a cylindrical pipe or plane channel flows.  However such simplifying procedures cannot be employed 
for non-spherical geometries (e.g., a duct). This is a second primary motivation for the present work.

Several strategies have been adopted by prior studies to incorporate the effect of the exterior magnetic field. One of them is the vertical field or pseudo-vacuum boundary
condition that has been used in several instances of astrophysical and dynamo simulations (\cite{Hurlburt:1988,Rüdiger:2001})
particularly due to its simplicity. An alternative method that was used in the simulation of the Karlsruhe dynamo experiment (see \cite{Rädler:1998,Rädler:2002}), was to immerse 
the conducting dynamo domain into a sphere, with the region between the sphere and the boundary of the conducting domain assumed to be filled with a material of low conductivity. 
However, both of these methods are associated with loss of solution accuracy. A rather straightforward procedure is to find a solution for the magnetic field in the exterior domain 
together with the interior \cite{Stefani:1999,Kenjeres:2006}. An approach similar to this but using the finite element method was proposed by Guermond et al.
\cite{Guermond:2003,Guermond:2007} and subsequently applied for dynamo problems (see \cite{Guermond:2009,Nore:2011}). This approach is however computationally demanding and is 
necessary only if one is interested in the solution of the exterior magnetic field. 

An alternative and elegant formulation, the velocity-current formulation, was first proposed and 
rigorously analyzed by Meir et al. \cite{Meir:1994, Meir:1996, Meir:1999} for stationary MHD flows and was further extended to time dependent flows by Schmidt \cite{Schmidt:1999}. This 
formulation takes advantage of the fact that the current density field is bounded within the domain (unlike the magnetic field) and instead of the induction equation for the magnetic field, an integro-differential 
transport equation for the current density is proposed. Subsequently Stefani et al. \cite{Stefani:2000,Xu:PRE-2004} introduced similar formuations (the integral equation approach) to
 kinematic dynamo problems and used it to simulate the 
von K\'{a}rm\'{a}n Sodium and Riga dynamo
experiments \cite{Xu:2008}. More recent developments and applications of this method can be found in \cite{Stefani:2013}. Nevertheless, it has been suggested that this procedure too 
requires large computational resources \cite{Giesecke:2008}, primarily due to the volume integrals that are required to be evaluated at every time step. Computationally more 
efficient is the coupled finite element-boundary integral approach that has been traditionally used to solve pure electromagnetic problems 
(see for e.g. \cite{Bossavit:1982,Bossavit:1991}). A finite-volume variant of this method was first proposed by Iskakov et al. \cite{Iskakov:2004,Iskakov:2005} 
to solve the induction equation and subsequently applied to kinematic dynamo simulations by Giesecke et al. \cite{Giesecke:2008}.

To our knowledge, DNS of MHD duct flow at finite $R_{m}$ with consistent treatment of the exterior domain has not been attempted in prior studies. In this paper, 
we apply the general approach of the coupled interior-exterior solution using the boundary integral procedure to the problem of turbulent magnetohydrodynamic flow in rectangular ducts. 
Specific geometric features such as the existence of corners and two non-periodic directions along with the need to treat magnetic diffusion in an implicit manner (unlike the case of 
high $R_{m}$ flows, where explicit schemes are typical) with integral boundary conditions, makes the problem computationally challenging. Here, we describe a divergence preserving 
semi-implicit hybrid finite-difference boundary integral numerical procedure for the problem of MHD duct flow with streamwise periodicity. As we will see in the last part, the differences
between our full MHD description and the quasistatic approximation of low-$R_{m}$ MHD can become significant, even for our DNS at moderate flow Reynolds numbers.

The paper is organized as follows. In Section 2 the physical model of the problem is introduced with the equations governing the interior and the exterior domains. In Section 3, 
the general details of the numerical procedure adopted for the hydrodynamic part is briefly described followed by the elaboration of boundary intergal approach and the algorithm 
for the coupled numerical procedure adopted to solve for the magnetic field. In Section 4 several test cases are presented in the limiting regime of low $R_{m}$ to 
verify the numerical implementation of the magnetic boundary conditions. Finally, in Section 5 the computational procedure is applied to perform DNS of turbulent
Hartmann duct flow to study the differences from the low -$R_{m}$ approximation arising in the evolution of turbulence at moderate $R_{m}$.

\section{Physical model and governing equations}
\begin{figure}[h]
\centerline{
\includegraphics[scale=0.36]{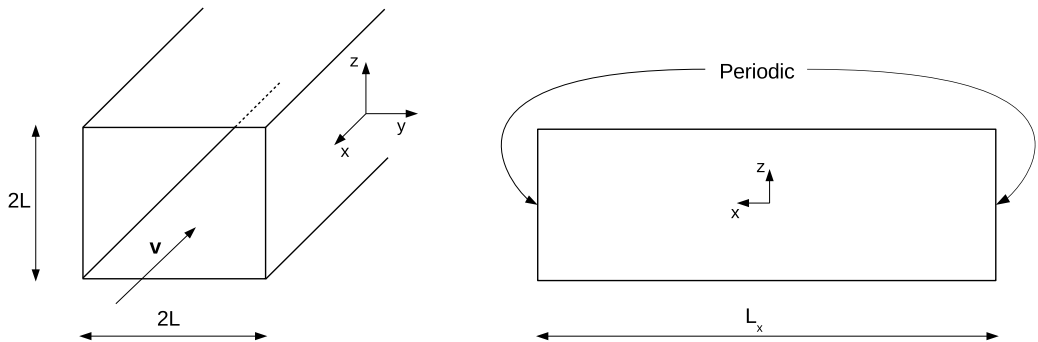}
}
\caption{Schematic of the flow in a straight rectangular duct with periodic inflow and outflow. Throughout this study $L_{y}=L_{z}=2L$.} 
\label{fig:duct}
\end{figure}

 We consider the turbulent flow of an incompressible and electrically conducting fluid (e.g., liquid metal or plasma) that is driven by a mean streamwise pressure gradient along a 
straight rectangular duct and is subjected to an externally imposed magnetic field $\bm{b}_{0}(\bm{x},t)$ (see Fig.~\ref{fig:duct}). The flow crossing the imposed magnetic field 
lines induces eddy currents $\bm{j}(\bm{x},t)$ in the fluid, which in turn produce a secondary (or induced) magnetic field $\bm{b}(\bm{x},t)$. The resultant magnetic field 
$\bm{b}_{t}=\bm{b}_{0}+\bm{b}$ interacts with the eddy currents to produce a Lorentz force that is proportional to $\bm{j}\times\bm{b}_{t}$ which affects the flow field. We 
are interested in the computation of the velocity and the magnetic fields in the interior of the duct through DNS. This means that the smallest 
scales, the Kolmogorov length and magnetic diffusion scales are resolved. Further, the mass flux through the duct is assumed to be constant and the direction along the mean 
flow (streamwise direction) is assumed to be periodic. 

In the case of a flow at low magnetic Reynolds number ($R_{m}\ll1$), the secondary magnetic field is assumed negligible when compared to the imposed magetic field and hence the evolution of 
such MHD flows can be described by the so called quasistatic or inductionless approximation \cite{Roberts-book}. However, when $R_{m}\sim1$, the case that we consider, the induced magnetic 
field is comparable to the imposed magnetic field and it becomes necessary to model the time evolution of $\bm{b}$. The physics of the coupled evolution of the flow and magnetic fields 
is described by the Navier-Stokes equations and the magnetic field transport equations respectively, together with solenoidal constraints for both fields. We denote the half-channel 
width as $L$, the average streamwise velocity as $U$ and the maximum value of the imposed magnetic field (generated from electric currents in the exterior)  on the duct walls as $B_{0}$. Upon non-dimensionalization with the scales $L$, $U$, $L/U$, $\rho U^2$, $B_{0}$ and $\sigma U B_{0}$ as the scales for the length, velocity, time, pressure, magneic field and current density respectively, and retaining the variable names for the non-dimensional quantities, the system of governing equations in the interior of the duct $\Omega_{i}$ can be written as

\begin{eqnarray}
\label{navierstokes} \hskip-15mm& & \frac{\partial \bm{v}}{\partial t} +
(\bm{v}\cdot \nabla)\bm{v} = -\nabla p +  \frac{1}{Re}\left( \nabla^2
\bm{v} + {\Har}^2\left(\bm{j} \times \bm{b}_{t}
\right)\right) ,\no{MHDeqn}\\
\label{btransport} \hskip-15mm& & \frac{\partial \bm{b}}{\partial t} +
(\bm{v}\cdot \nabla)\bm{b}_{t} = (\bm{b}_{t}\cdot \nabla)\bm{v} + \frac{1}{R_{m}}\nabla^2
\bm{b},\no{MHDeqn}\\
\label{continuity} \hskip-15mm& & \nabla \cdot\bm{v}  =  0,\\
\label{divb} \hskip-15mm& & \nabla \cdot\bm{b}  =  0,\\
\label{amperelaw} \hskip-15mm& & \bm{j} = \frac{1}{R_{m}}\left( \nabla \times\bm{b}\right) ,\\
\label{MHDeqn_bc} \hskip-15mm& & u = v = w =0 \hskip2mm\mbox{on } \varSigma \hskip1mm; \quad
\textrm{$\bm{v}$, $\bm{b}$ periodic in $x$-direction} \no{MHDeqn_bc}
\end{eqnarray}
where $x$, $y$ and $z$ denote the streamwise, spanwise and wall normal directions respectively. The standard no-slip and no penetration boundary conditions are assumed for $\bm{v}$ on the wall boundaries $\varSigma$, along with periodicity in the streamwise direction. The duct walls $\varSigma$ are considered to be electrically insulating ($\sigma=0$ on $\varSigma$) which translates to vanishing wall normal current density $j_{n}=0$. 

The non-dimensional parameters involved in the system are the Reynolds number ($\Rey$), the Hartmann number ($\Har$) defined as
\begin{equation}
\label{non-dim} \Rey= \frac{UL}{\nu} \textrm{,}\hskip7mm \Har= B_{0}L\sqrt{\frac{\sigma}{\rho\nu}}
\end{equation}
and the magnetic Reynolds number ($R_{m}$). Here, $\nu$ and $\rho$ represent the kinematic viscosity and the density of the fluid respectively. The magnetic Prandtl number relates the magnetic and hydrodynamic Reynolds numbers, and is defined as
\begin{equation}
Pr_{m}=\frac{R_{m}}{\Rey} \textrm{.}
\end{equation}
However, we treat $R_{m}$ as an independent parameter (instead of $Pr_{m}$) throughout this paper.

The region outside the duct $\Omega_{e}$ is considered to be electrically insulating (e.g., air or vacuum). It is evident that the secondary magnetic field is not limited to the duct interior but extends across the duct walls and pervades the space outside the duct. This happens unless the duct walls are perfectly conducting, in which case the magnetic field is bound within the interior of the duct. We denote the secondary magnetic field extending outside the duct as the exterior magnetic field. Although our primary interest is in the magnetic field inside the duct, a consistent treatment requires that the magnetic field is continuous across the duct walls. This is ensured by considering the magnetic field in the extended domain including the region outside the duct. Since electric currents cannot exist in the exterior, the magnetic field is curl free and hence can be expressed as the gradient of a magnetic scalar potential, $\bm{b}=-\nabla\psi$. Imposing the solenoidality condition (Gauss law, $\nabla\cdot\bm{b}=0$) yields the following governing equations for the magnetic field in the exterior
\begin{eqnarray}
\label{laplace} \hskip-15mm& & \nabla^2\psi = 0 \hskip1mm \textrm{,}\hskip2mm \bm{b}=-\nabla\psi \hskip3mm \textrm{in} \hskip1mm \Omega_{e}\cup\varSigma \textrm{,} %\\ 
%\label{farfield} \hskip-15mm& & \psi\rightarrow\mathcal{O}(r^{-2}) \textrm{,} \hskip2mm r\rightarrow\infty
\end{eqnarray}
where $\varSigma$ represents the duct wall boundary. In addition, it is assumed that no net streamwise current is applied, due to which the scalar potential $\psi$ far away from the walls 
decreases faster than $\mathcal{O}(r^{-1})$ as $r\rightarrow\infty$, satisfying the far field condition, where $r$ is the normal distance from the ducts walls.
 Equations \eqref{navierstokes} to \eqref{MHDeqn_bc} together with \eqref{laplace} and the far field condition completely determine the physical system under consideration. 
However, since we are interested only in the solution of the magnetic field inside the duct, by means of the boundary integral approach, boundary conditions are obtained for the magnetic field that characterizes the matching of the exterior and interior fields at the wall boundary. This leads to non-local magnetic boundary conditions on the duct walls. A detailed discussion of the boundary integral procedure and the particular form of the non-local conditions will follow in a later section of this paper.

\section{Numerical procedure}

\subsection{The interior problem}
The governing partial differential equations for the velocity and the magnetic fields inside the duct are solved numerically using the finite differences approach. The domain is discretized 
into a structured rectangular Cartesian grid and the solution variables are approximated at the grid points which correspond to the collocated grid arrangement. In duct MHD flows in a 
uniform external magnetic field, specific boundary layers with steep velocity gradients and high current densities are formed near the walls \cite{Muller-book}. These correspond to the 
Hartmann layers at the walls normal to the imposed magnetic field and the Shercliff layers at the walls aligned to the initial magnetic field $\bm{b}_{0}$. In order to resolve the thin boundary layers, the grid in the cross section 
is stretched to obtain a non-uniform grid with high grid clustering near the walls. The non-uniform grid in both wall-normal directions is obtained by a coordinate transformation from 
the uniform-grid coordinates $\left( \zeta,\eta \right)$ according to
\begin{equation}
\label{eq:gridstretch} y = L\frac{\tanh(S_{y}\zeta)}{\tanh(S_{y})} \hskip2mm \textrm{,  } \hskip2mm z = L\frac{\tanh(S_{z}\eta)}{\tanh(S_{z})} \hskip2mm \textrm{,}
\end{equation}
where $S_{y}, S_{z} $ correspond to the degree of stretching in the $y$ and $z$ directions respectively. However, a uniform grid in the $x$-direction is considered so as to take advantage of the periodicity through Fourier decomposition.

% Insert a picture of stencil showing the location of variable life in the grid
In order to keep the paper self contained, we now briefly describe the computational procedure adopted for the solution of velocity field from the Navier-Stokes equations. The time discretization is performed by a second-order backward difference scheme using the 3 time levels $n-1$, $n$, $n+1$ when marching from time level $n$ to $n+1$ as
\begin{equation}
\frac{\partial \bm{v}}{\partial t} \approx \frac{3{\bm{v}}^{n+1}-4{\bm{v}}^{n} + {\bm{v}}^{n-1}}{2\Delta{t}} \textrm{.}
\end{equation}   
The viscous term can be treated using either an explicit or implicit procedure, whereas the non-linear advective term and the Lorentz force term are treated explicitly using the Adams-Bashforth method. The advective, Lorentz force and viscous terms can be summed up into ${\bm{F}}^n$ as 
\begin{equation}
{\bm{F}}^n = -\left( {\bm{v}}^n\cdot\nabla\right){\bm{v}}^n + \frac{\Har^2}{\Rey}\left( \bm{j}\times {\bm{b}_{t}}^n\right) +  \frac{\left(1-\theta \right)}{\Rey}\nabla^2{{\bm{v}}^n} \textrm{,}
\end{equation} 
where binary factor $\theta$ assumes the values $0$ and $1$ for the explicit and implicit treatments respectively. The implicit treatment of the viscous term can be advantageous for the case of small $\Rey$. The velocity field is obtained by the well known projection method, wherein an intermediate velocity field ${\bm{v}}^*$ is computed using
\begin{equation}
\label{velocitypoisson} \frac{3{\bm{v}}^*-4{\bm{v}}^{n} + {\bm{v}}^{n-1}}{2\Delta{t}} = 2{\bm{F}}^n - {\bm{F}}^{n-1} +  \frac{\theta}{\Rey}\nabla^2{{\bm{v}}^*} \textrm{,}
\end{equation} 
which leads to a Poisson-type equation for ${\bm{v}}^*$ in the implicit case.
The pressure field $p^{n+1}$ is then computed from the continuity equation by solving another Poisson problem, 
\begin{equation}
\label{pressurepoisson} \nabla^2{p^{n+1}} = \frac{3}{2\Delta{t}}\nabla \cdot\bm{v^*} \textrm{.}
\end{equation} 
Integrating \eqref{pressurepoisson} over the whole domain and applying the Gauss-divergence theorem will yield the boundary condition for pressure on $\varSigma$ as
\begin{equation}
\frac{\partial p^{n+1}}{\partial n} = \frac{3}{2\Delta{t}} {\bm{v}}_{nor}^{*} \textrm{,}
\end{equation} 
where the subscript $nor$ refers to the wall normal component. Subsequently the intermediate non-solenoidal velocity field ${\bm{v}}^*$ is projected onto a divergence-free velocity
 field $\bm{v}$ at the time level $n+1$ using the pressure field obtained from \eqref{pressurepoisson} as
\begin{equation}
{\bm{v}}^{n+1}={\bm{v}}^{*} - \frac{2\Delta{t}}{3} \nabla{{p}^{n+1}} \textrm{.}
\end{equation} 
A Fourier transformation is applied  in the $x$-direction to the discrete forms of the Poisson equations \eqref{velocitypoisson}, \eqref{pressurepoisson} for the velocity and pressure. The transformed equations are then solved in the wavenumber space as a series of 2D ($y,z$ plane) problems using the Fortran software package FISHPACK \cite{FISHPACK} that uses a cyclic reduction algorithm (direct solver) for the solution of 2D elliptic equations. Further details of the numerical procedure for the hydrodynamic solution can be found in Krasnov et al. \cite{Krasnov:2011}.

We now turn our attention to the solution of the magnetic induction equation \eqref{btransport}. A discretization procedure similar to that used for the implicit treatment of the momentum equation is followed with only the diffusive term treated implicitly. Unlike the momemtum equation, the implicit treatment here is really essential due to the fact that the diffusive time scale in the case of $R_{m}\sim1$ is comparable to the time scale of advection of the magnetic field. Discretization of the induction equation yields 
% Discretized induction equation
\begin{equation}
\label{adamsbashforth} \frac{3{{\bm{b}}^{n+1}}-4{\bm{b}}^{n} + {\bm{b}}^{n-1}}{2\Delta{t}} = 2{\bm{T}}^n - {\bm{T}}^{n-1} + \frac{1}{R_{m}}\nabla^2{{\bm{b}}^{n+1}}
\end{equation}
for the secondary magnetic field $\bm{b}$ at the n+1 level where ${\bm{T}}^n$ includes the advective and the magnetic field stretching terms and is given by
\begin{equation}
{\bm{T}}^n = -\left( {\bm{v}}^n\cdot\nabla\right){\bm{b}_{t}}^n + \left( {\bm{b}_{t}}^n\cdot\nabla\right){\bm{v}}^n \textrm{.}
\end{equation} 
Further simplification of \eqref{adamsbashforth} leads to a Poisson-type equation for ${\bm{b}}^{n+1}$ as
\begin{equation}
\label{bpoisson} -f \bm{b}^{n+1} + \nabla^2\bm{b}^{n+1}  = -f\bm{q} \textrm{,}
\end{equation} 
where $f=\frac{3R_{m}}{2\Delta{t}}$ is a discretization coefficient and $\bm{q}$ is the right hand side that retains the known terms from the time steps $n$ and $n-1$. The system being periodic in the streamwise direction, we now introduce a Fourier transformation in the $x$-direction as
\begin{equation}
\label{fft} \bm{b}(x,y,z) = \Re \left\lbrace \sum\limits_{k=0}^{k=\frac{N_{x}}{2}-1}\bm{\hat{b}}_{k}(y,z) e^{i\alpha_{k}x} \right\rbrace  \textrm{,}
\end{equation}
where $\Re$ represents the real part, $N_{x}$ is the number of grid intervals along the $x$-direction and $\alpha_{k}$ is the streamwise wavenumber defined as $\alpha_{k}=2\pi k/L_{x}$, $L_{x}$ being the length of the duct. Substituting \eqref{fft} into \eqref{bpoisson} leads to a 2D elliptic equation in the $yz$-plane for the complex Fourier coefficients $\bm{\hat{b}}_{k}$ as 
\begin{equation}
\label{bpoissonfspace} \left( -f -\alpha_{k}^2 \right)\bm{\hat{b}}_{k} + \nabla^2_{yz}\bm{\hat{b}}_{k}  = -f\bm{\hat{q}}_{k} \textrm{,}
\end{equation}
with $\bm{\hat{b}}_{k}=\left[  \hat{b}_{xk}, \hat{b}_{yk}, \hat{b}_{zk} \right]$ and $\nabla^2_{yz}=\partial_{y}^2 + \partial_{z}^2$ is the 2D Laplace operator. This step is essential as we reduce the complexity of matching the magnetic field of a three-dimensional interior ($\Omega_{i}$) and an exterior ($\Omega_{e}$) to a planar problem for each Fourier coefficient. Here, the superscript $n+1$ is dropped for the sake of simplicity. Solution of \eqref{bpoissonfspace} requires proper boundary conditions for the magnetic field that matches the exterior field, which will be the subject of the following section.

\subsection{Boundary integral equation and the coupled numerical procedure}
In this section we will derive suitable boundary conditions (in the Fourier space) required for the closure of \eqref{bpoissonfspace} and present a coupled iterative solution procedure to solve the resulting system. This is done through the boundary integral approach, by which the matching of the interior solution with the exterior solution at the boundary translates into non-local boundary conditions. The governing Laplace equation \eqref{laplace} for the exterior magnetic potential transforms to the 2D Helmholtz equation in the $k$-space as
% Helmholtz equation and the Boundary Integral equations
\begin{equation}
\label{helmholtz} (\nabla^2_{yz}-\alpha_{k}^2){\hat{\psi}}_{k} =  0 \textrm{.}
\end{equation}

\begin{figure}[h]
\centerline{
\includegraphics[scale=0.55]{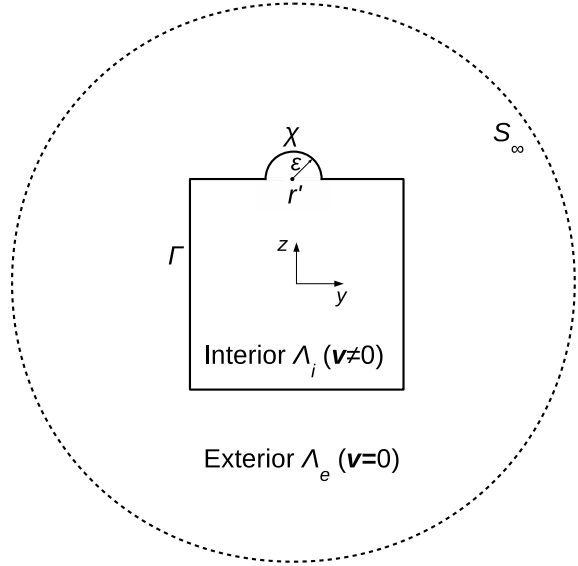}
}
\caption{Region of integration between the rectangular boundary $\varGamma$ and an outer circle $S_{\infty}$ excluding a small semi-circle $\chi$ of radius $\varepsilon$. This part of the solution procedure is done in a plane. Thus $\varLambda_{i}$ relates to $\Omega_{i}$, $\varLambda_{e}$ to $\Omega_{e}$ and $\varGamma$ to $\varSigma$ from the original 3D setting.} 
\label{fig:bigball}
\end{figure}

The Green's function or the fundamental solution of the 2D Helmholtz operator is denoted by $G_{k}\left( \bm{r'},\bm{r}\right) $ that satisfies $(\nabla^2_{yz}-\alpha_{k}^2)G_{k}(\bm{r'},\bm{r}) = -\delta(\bm{r'}-\bm{r})$ where $\delta(\bm{r'}-\bm{r})$ is the Dirac delta function centered around the pole $\bm{r'}=y'\bm{j}+z'\bm{k}$, with $\bm{j}$ and $\bm{k}$ representing the unit vectors in the $y$ and $z$ directions respectively (see Fig. \ref{fig:bigball}). Considering $\bm{r'}$ to be a point on the rectangular boundary $\varGamma$, we see that

\begin{align}
\label{greenssecond} 
\begin{split}
&\int_{\varLambda_{e}}\nabla\cdot\left( {\hat{\psi}}_{k}\left( \bm{r}\right) \nabla G_{k}(\bm{r'},\bm{r}) - G_{k}(\bm{r'},\bm{r}) \nabla {\hat{\psi}}_{k}\left( \bm{r}\right)\right) dA \\&= \int_{\varLambda_{e}}\left({\hat{\psi}}_{k}\left( \bm{r}\right) \nabla^2 G_{k}(\bm{r'},\bm{r}) - G_{k}(\bm{r'},\bm{r})\nabla^2 {\hat{\psi}}_{k}\left( \bm{r}\right)\right) dA \\
&= \int_{\varLambda_{e}}{\hat{\psi}}_{k}\left( \bm{r}\right) \delta\left( \bm{r'}-\bm{r}\right) dA = 0 \textrm{,}
%{\Omega_{e}}
\end{split}
\end{align}
where the area of integration includes the exterior region between the big circle and the rectangular domain excluding a small semi-circle of radius $\varepsilon$ in the vicinity of the pole $\bm{r'}$ as shown in Fig. \ref{fig:bigball}.

Using Gauss-divergence theorem, equation \eqref{greenssecond} can be rewritten as
%\begin{multline*}
\begin{align}
\begin{split}
&\int\displaylimits_{\varGamma}\left( {\hat{\psi}}_{k}\left( \bm{r}\right)\nabla G_{k}(\bm{r'},\bm{r}) - G_{k}(\bm{r'},\bm{r}) \nabla {\hat{\psi}}_{k}\left( \bm{r}\right)\right)\cdot\bm{n} dl \\ &+ \int\displaylimits_{\chi}\left( {\hat{\psi}}_{k}\left( \bm{r}\right)\nabla G_{k}(\bm{r'},\bm{r}) - G_{k}(\bm{r'},\bm{r}) \nabla {\hat{\psi}}_{k}\left( \bm{r}\right)\right)\cdot\bm{n} dl \\
 &+ \int\displaylimits_{S_{\infty}}\left( {\hat{\psi}}_{k}\left( \bm{r}\right)\nabla G_{k}(\bm{r'},\bm{r}) - G_{k}(\bm{r'},\bm{r}) \nabla {\hat{\psi}}_{k}\left( \bm{r}\right)\right)\cdot\bm{n} dl= 0 \textrm{,}
\end{split}
\end{align}
%\end{multline*}
where $\bm{n}$ is the local unit outward normal vector at $\bm{r}$ on the boundaries and $l$ is the arc length. The third term (integral over $S_{\infty}$) vanishes as $\bm{r}\rightarrow\infty$ and the second term (integral over $\chi$) is simplified with the assumption that ${\hat{\psi}}_{k}$ and $\frac{\partial {\hat{\psi}}_{k}}{\partial n}$ do not vary within the half-circle $\chi$ as the radius $\varepsilon$ is considered small.  

%\begin{multline*}
\begin{equation}
\begin{split}
\int\displaylimits_{\chi} \hat{\psi}_{k}\left( \bm{r}\right) \nabla G_{k}(\bm{r'},\bm{r}) \cdot\bm{n} dl &= \hat{\psi}_{k}\left( \bm{r'}\right) \int\displaylimits_{S}\nabla^2 G_{k}(\bm{r'},\bm{r})dS \\&= \hat{\psi}_{k}\left( \bm{r'}\right) \int\displaylimits_{S} \left( -\delta\left( \bm{r'}-\bm{r}\right) + \alpha^2 G_{k}(\bm{r'},\bm{r})\right)  dS \\
&= -\frac{1}{2} \left(\bm{r'}\right)\hat{\psi}_{k}\left( \bm{r'}\right) +  \alpha^2 \hat{\psi}_{k}\left( \bm{r'}\right) \int\displaylimits_{S}G_{k}(\bm{r'},\bm{r}) dS  \textrm{,} 
\end{split}
\end{equation}
%\end{multline*}
where $l$ is the coordinate along $\chi$ and $S$ represents the area encumbered by the half-circle $\chi$. Using the fact that $G_{k}(\bm{r'},\bm{r}) \sim \ln(\lvert \bm{r'}-\bm{r}\rvert)$ for small $\varepsilon$, the second term in the above equation vanishes as $\varepsilon\rightarrow 0$. In the case when $\bm{r'}$ lies at one of the four corners of $\varGamma$, $\chi$ would correspond to a three-quarter circle. Furthermore,

\begin{equation}
\begin{split}
\int\displaylimits_{\chi}-G_{k}(\bm{r'},\bm{r}) \nabla \hat{\psi}_{k}\left( \bm{r}\right)\cdot\bm{n} dl &= -\frac{\partial {\hat{\psi}}_{k}}{\partial n}\left( \bm{r'}\right) \int\displaylimits_{\chi} G_{k}(\bm{r'},\bm{r}) dl \\ &= -\frac{\partial {\hat{\psi}}_{k}}{\partial n}\left( \bm{r'}\right) \ln \left( \varepsilon\right)  2\pi\varepsilon \rightarrow 0, \hskip3mm \varepsilon\rightarrow 0  \textrm{.}
\end{split}
\end{equation}
With the above simplifications, the boundary integral equation in the general form can be written as
\begin{equation}
\label{bie} \beta\left(\bm{r'}\right) \hat{\psi}_{k}(\bm{r'}) =  \text{P.V.}\oint_{\varGamma}[G_{k}(\bm{r'},\bm{r})\hat{b}_{nk}(\bm{r}) + \hat{\psi}_{k}(\bm{r})\frac{\partial G_{k}}{\partial n}(\bm{r'},\bm{r})]dl(\bm{r})  \textrm{,}
\end{equation}
where $\hat{b}_{nk}(\bm{r})=-\frac{\partial {\hat{\psi}}_{k}}{\partial n}(\bm{r})$, $\beta\left( \bm{r'}\right)$ is a constant that depends on the location of the pole $\bm{r'}$ on the rectangular boundary $\varGamma$ and is given by
\begin{eqnarray}
\beta\left(\bm{r'}\right) = 
\begin{cases} 
   \frac{3}{4},& \text{if } \bm{r'}\in \textrm{corner} \\
   \frac{1}{2},& \textrm{otherwise}  \textrm{,}
\end{cases}
\end{eqnarray}
$n$ being the local outward wall normal coordinate at $\bm{r}$. It should be mentioned that the integration along the rectangular contour $\varGamma$ must be performed in the sense of a Cauchy principal value (CPV) \cite{Bronshtein-book}. The boundary condition \eqref{bie} is a Fredholm integral equation of the $2^{\text{nd}}$ kind with a singular kernel. The singularity would be apparent from the specific form of the Green's function given by  
% Particular forms of the 2D Green's functions 
\begin{equation}
G_{k}\left(\bm{r'},\bm{r} \right)  = \frac{1}{2\pi}K_{0}\left( \alpha_{k}\left( \lvert \bm{r'}-\bm{r}\rvert\right) \right) \textrm{,}  
\end{equation}
$K_{0}$ being the MacDonald function which corresponds to the complex valued Hankel function of zero order $H_{0}$ \cite{Stakgold-book}. For numerical evaluation, the following series expansion formulae from \cite{Abramowitz-book} are particularly useful 

\begin{eqnarray}
K_{0}\left(x\right) = 
\begin{cases}
    -\ln{\left( \frac{x}{2} \right)} I_{0}\left(x\right) + \sum\limits_{n=1}^7{C_{n}\left(\frac{x^2}{4}\right)^{n-1}},& \text{if } x\leq2 \\
    \frac{e^{-x}}{\sqrt{x}}\sum\limits_{n=1}^7{D_{n}\left(\frac{2}{x}\right)^{n-1}}, & \text{otherwise} 
\end{cases} \\
I_{0}\left(x\right)  = 1 + \sum\limits_{n=1}^6{E_{n}y^{n}} \textrm{,} \hskip9mm y = {\left( \frac{x}{3.75}\right) }^2 \textrm{,} \hskip9mm \lvert x\rvert<3.75 \textrm{,}
\end{eqnarray}
in which $I_{0}$ represents the modified Bessel function of the first kind and $C_{n},D_{n},E_{n}$ are the series coefficients \cite{Abramowitz-book}. It can be seen that for $x\rightarrow0$, $K_{0}\left(x\right)\sim-\ln{x}$ which explains the logarithmic singularity at the pole. 

Solution of \eqref{bpoissonfspace} for the in-plane components ${\hat{b}}_{yk}$ and ${\hat{b}}_{zk}$ requires the normal and tangential components ${\hat{b}}_{n}$ and ${\hat{b}}_{\tau}$ on the boundary $\varGamma$ which are connected through the potential ${\hat{\psi}}_{k}$ given by \eqref{bie}. The normal component $\hat{b}_{n}$ on the boundary can be evaluated from the Gauss law as
% Equations pertaining to the coupled-BEM solution
\begin{equation}
\label{bneqn} \frac{\partial \hat{b}_{nk}}{\partial n} + \frac{\partial \hat{b}_{\tau k}}{\partial \tau} = -\alpha_{k}^2 \hat{\psi}_{k}
\end{equation}
and the tangential component $\hat{b}_{\tau k}$ obtained from
\begin{equation}
\label{bteqn} \hat{b}_{\tau k} = -\frac{\partial \hat{\psi}_{k}}{\partial \tau} \textrm{,}
\end{equation}
which closes the problem of evaluating the in-plane components ${\hat{b}}_{yk}$ and ${\hat{b}}_{zk}$. 

Equations \eqref{bpoissonfspace}, \eqref{bneqn} and \eqref{bteqn} are discretized by finite differences and equation \eqref{bie} is discretized by the boundary element method and are solved together iteratively for the numerical solution of the two components. A coupled iterative procedure between the interior and the boundary has been adopted here. The discrete form of the elliptic equation \eqref{bpoissonfspace} is used to update ${\hat{b}}_{yk}$ and ${\hat{b}}_{zk}$ in the strict interior by a Gauss-Seidel like method using boundary values from the previous iteration. The component of ${\hat{b}}_{nk}$ on grid points adjacent to the boundary is then used to update ${\hat{b}}_{nk}$ on $\varGamma$ through \eqref{bneqn}. The updated ${\hat{b}}_{nk}$ is used to update $\hat{\psi}_{k}\left( \bm{r'}\right) $ on $\varGamma$ through the discrete form of \eqref{bie} which is subsequently used to evaluate ${\hat{b}}_{\tau k}$ from \eqref{bteqn}. This iterative procedure alternating between the interior and the boundary is performed until the required convergence criterion is met. The procedure for a single iteration is summarized below

\begin{itemize}
  \item Compute $\hat{b}_{yk}$ and $\hat{b}_{zk}$ on $\varLambda_{i}$ with $\hskip2mm \left( -f -\alpha_{k}^2 \right)\bm{\hat{b}}_{k} + \nabla^2_{yz}\bm{\hat{b}}_{k}  = -f\bm{\hat{q}}_{k}$ 
  \item Compute $\hat{b}_{nk}$ on $\varGamma$ with $\hskip2mm  \frac{\partial \hat{b}_{nk}}{\partial n} + \frac{\partial \hat{b}_{\tau k}}{\partial \tau} = -\alpha_{k}^2 \hat{\psi}_{k}$
  \item Compute $\hat{\psi}_{k}$ on $\varGamma$ with \\ $\hskip2mm  \beta\left(\bm{r'}\right) \hat{\psi}_{k}(\bm{r'})  
   =\text{P.V.}\oint_{\varGamma}[G_{k}(\bm{r'},\bm{r})\hat{b}_{nk}(\bm{r}) + \hat{\psi}_{k}(\bm{r})\frac{\partial G_{k}}{\partial n}(\bm{r'},\bm{r})]dl(\bm{r})$
  \item Compute $\hat{b}_{\tau k}$ on $\varGamma$ with $\hskip2mm \hat{b}_{\tau k} = -\frac{\partial \hat{\psi}_{k}}{\partial \tau}$ \textrm{.}
\end{itemize}  
It must be noted that although it is possible to use direct solvers to solve the discretized forms of equations \eqref{bpoissonfspace}, \eqref{bneqn}, \eqref{bteqn} and \eqref{bie}, the iterative procedure is found to be computationally
efficient mainly due to the very good initital guess obtained for the unknown variables from the previous time step.

We now turn to the discretization of the boundary integral equation \eqref{bie} which forms the basis of the coupled iterative procedure just described. 
\begin{figure}[h]
\centerline{
\includegraphics[scale=0.55]{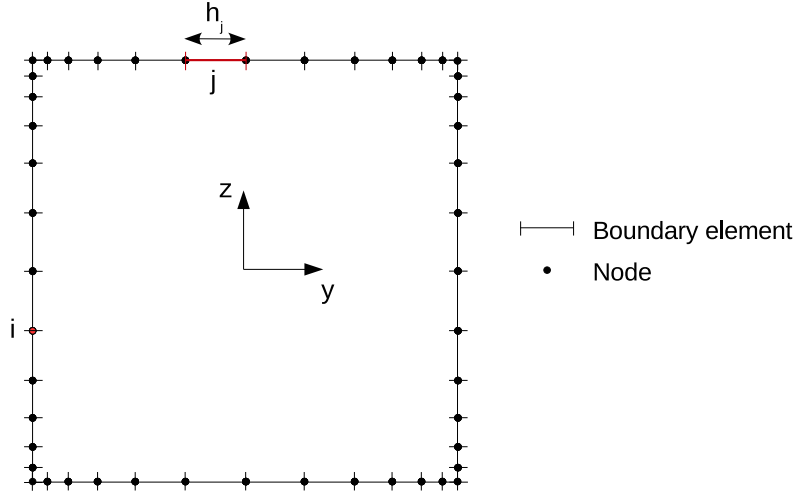}
}
\caption{Representative discretization of the rectangular boundary $\varGamma$ into nodes and boundary elements.} 
\label{fig:belements}
\end{figure}
Equation \eqref{bie} is discretized to obtain a set of algebraic equations by the formalism of boundary element method \cite{Brebbia-book}. The rectangular boundary is divided into a number of small line segments called boundary elements and the contour integral along $\varGamma$ is approximated as a sum of integrals along each of these elements. The solution variable $\hat{\psi}_{k}$ is approximated at the ends of the boundary elements which are denoted as nodes. The location of the boundary elements and nodes are shown in Fig.~\ref{fig:belements}. This layout of the elements leads to a double node at each of the four corners of $\varGamma$ which is essential in order to deal with the singularity that exists for the normal vector $\hat{\bm{n}}$ at the corners. A piecewise linear variation of $\hat{\psi}_{k}$ is assumed along each element. Denoting the length of the $j^{\text{th}}$ element by $h_{j}$ and temporarily omitting the subscript $k$ for simplicity of notation, the discrete version of \eqref{bie} for node $i$ at $\bm{r'}$ can then be written as

\begin{equation}
\begin{split}
% Discrete equation for BIE
\label{discretebie} &\beta_{i}\hat{\psi}_{i} -  \sum\limits_{j=1}^{j=N_{b}} \int\displaylimits_{0}^{h_{j}} \left( \frac{{\hat{\psi}}_{nj}\left(h_{j}-l \right) +  {\hat{\psi}}_{nj+1} l}{h_{j}}\right)\frac{\partial G}{\partial n} (\bm{r'}_{i},\bm{r}_{j})dl   \\ &=  \sum\limits_{j=1}^{j=N_{b}} \int\displaylimits_{0}^{h_{j}} \left( \frac{{\hat{b}}_{nj}\left(h_{j}-l \right) +  {\hat{b}}_{nj+1} l}{h_{j}}\right) G\left( \bm{r'}_{i},\bm{r}_{j}\right) dl
\end{split}
\end{equation}
for $1\leq i \leq N_{b}$, $N_{b}$ being the number of boundary nodes (see Fig.~\ref{fig:belements}). The numerical integral on the right hand side is evaluated along each element using a 4-point Gauss-Legendre quadrature \cite{Abramowitz-book}. This is important in order to be able to capture the steep gradients in the Green's function for the wide range of wavenumbers ($\alpha_{k}$) involved. Caution is necessary for the computation of the integral in the region containing the pole $\bm{r'}_{i}$ as the function $K_{0}$ is singular at the pole. The logarithm poses a weak singularity and is dealt by analytical integration over the two elements lying on either side of the node $i$ which is possible since the integral is convergent \cite{Christiansen:1971}. 

Through this procedure we obtain a linear system of equations for $\hat{\psi}_{k}$ as 
\begin{equation}
% Equation A psi = b
S\underline{\hat{\psi}_{k}}=\underline{m} \textrm{.}
\end{equation}
The matrix $S$ is fully occupied due to the non-local nature of the boundary conditions and vector $\underline{m}$ contains the right hand side of \eqref{discretebie}. This concludes the numerical computation of the in-plane components ${\hat{b}}_{yk}$ and ${\hat{b}}_{zk}$, and it remains to evaluate the streamwise component ${\hat{b}}_{xk}$ which will be discussed next. 

In principle, the streamwise Fourier coefficient ${\hat{b}}_{xk}$ can be computed from the discrete form of the induction equation in $k$-space \eqref{bpoissonfspace}, with the Dirichlet condition ${\hat{b}}_{xk}=-i\alpha_{k}\hat{\psi}_{k}$ on the boundary $\varGamma$. However this raises the issue of preserving the divergence of the magnetic field ($\nabla\cdot\bm{b}=0$) during the course of its evolution, due to the reason that equations \eqref{btransport} and \eqref{divb} form an overdetermined system for the $\bm{b}$ field. Maintaining $\nabla \cdot{\bm{b}} = 0$ numerically is a non-trivial issue and various strategies are often adopted to ensure solenoidality (see \cite{Toth:JCP-2000} for a detailed discussion). The issue becomes even more challenging when a semi-implicit or a fully implicit procedure is used for the magnetic field along with non-local boundary conditions. The numerical source of generation of $\nabla\cdot\bm{b}$ can be understood as follows. Taking the divergence of \eqref{bpoisson} and rearranging the terms gives
% Transport equation for div.B
\begin{equation}
D^{n+1} = \frac{1}{f}\nabla^2 D^{n+1} + D_{q} \textrm{,}
\end{equation}
where $D^{n+1}=\nabla \cdot{\bm{b}}^{n+1}$ and $D_{q}=\nabla \cdot\bm{q}$. Although the initial fields $\bm{v}^n$ and $\bm{b}^n$ are divergence-free (hence the last term on the right hand side vanishes), the boundary conditions act as a source of $D^{n+1}$ during the solution of the Poisson equation for $D^{n+1}$. This contaminates $D^{n+1}$ on the interior points adjacent to the boundary and the divergence diffuses into the domain interior subsequently. In the appendix we explain using a simple example, the generation of $D^{n+1}$ on the interior grid points adjacent to the boundary due to the implicit treatment of the diffusion term in the induction equation.

In order to preserve the solenoidality of the magnetic field, the streamwise component $\hat{b}_{xk}$ is reconstructed from the in-plane components using
\begin{equation}
\hat{b}_{xk} = \frac{-1}{i\alpha_{k}}\left(\frac{\partial \hat{b}_{yk}}{\partial y} + \frac{\partial \hat{b}_{zk}}{\partial z}\right),\hskip2mm\textrm{for wavenumbers}\hskip1mm \:k\neq0 \textrm{.}
\end{equation}
This ensures a divergence-free magnetic field for all the non-zero Fourier modes.

\subsection{Treatment of the zero mode ($k=0$)}
The reconstruction of $b_{x}$ is however not possible for the zero mode due to the reason that when $k=0$, the streamwise mean component $\bar{b}_{x}$ is decoupled from the in-plane mean components $\bar{b}_{y}$ and $\bar{b}_{z}$, where the overbar denotes averaging with respect to $x$. Hence we solve \eqref{bpoissonfspace} for the mean component $\bar{b}_{x}$ which can be written as
\begin{equation}
\label{bxnpoissonfspace} -f\bar{b}_{x} + \nabla^2_{yz}\bar{b}_{x}  = -f\bar{q}_{x} \textrm{.}
\end{equation}
The boundary condition for this is obtained again from $\nabla\times\bar{\bm{b}}=0$ which leads to the Dirichlet condition $\bar{b}_{x}=constant$ and the constant can be conveniently chosen to be zero
% bx mean =0 equation
\begin{equation}
\bar{b}_{x}=0 \textrm{.}
\end{equation}
The discrete form of \eqref{bxnpoissonfspace} is solved with the Dirichlet BCs using the Poisson solver similar to that of pressure. 

Since reconstruction of $\bar{b}_{x}$ is not possible when $k=0$, satisfying $\nabla\cdot\bar{\bm{b}}=0$ is not guaranteed with the usage of primitive variables. Therefore the mean in-plane components $\bar{b}_{y}$ and $\bar{b}_{z}$ are computed through the magnetic vector potential $A$ which is defined by
% Equations for the k=0 mode
\begin{equation}
\label{adefinition} \bar{b}_{y}=\frac{\partial A}{\partial z} \textrm{,} \hskip9mm  \bar{b}_{z}=-\frac{\partial A}{\partial y} \textrm{.}
\end{equation}
The governing equation for $A$ is derived as follows. Averaging equation \eqref{btransport} along the $x$-direction and rewriting the advective and field stretching terms in the curl form gives 
% x-averaged b-transport equation
\begin{equation}
\frac{\partial \bar{\bm{b}}}{\partial t} = \overline{\nabla\times(\bm{v}\times\bm{b} )} + \frac{1}{R_{m}}\nabla^2_{yz}\bar{\bm{b}} \textrm{.}
\end{equation}
Further simplification yields the mean equations for the in-plane components as 
% mean-equation for by_ and bz_
\begin{eqnarray}
\hskip-15mm& & \frac{\partial \bar{b}_{y}}{\partial t} = \frac{\partial}{\partial z}\left(\overline{v_{y}b_{zt}-b_{y}v_{zt}} \right) + \frac{1}{R_{m}}\nabla^2_{yz}\bar{b}_{y} \textrm{,} \\
\hskip-15mm& & \frac{\partial \bar{b}_{z}}{\partial t} = -\frac{\partial}{\partial y}\left(\overline{v_{y}b_{zt}-b_{y}v_{zt}} \right) + \frac{1}{R_{m}}\nabla^2_{yz}\bar{b}_{z} \textrm{.}
\end{eqnarray}
Introducing the vector potential and integrating yields the following governing equation for $A$ in the interior
\begin{equation}
\label{atransport} \frac{\partial A}{\partial t} = \overline{v_{y}b_{zt}-b_{yt}v_{z}} + \frac{1}{R_{m}}\left( \frac{\partial^2 A}{\partial y^2} + \frac{\partial^2 A}{\partial z^2}\right) + \varsigma\left( t\right) \textrm{,}  
\end{equation}
where $\varsigma\left( t\right) $ is a constant of integration that depends only on time.

In the exterior, $\overline{\nabla\times\bm{b}}=0$ yields 
\begin{equation}
\label{laplacea} \frac{\partial^2 A}{\partial y^2} + \frac{\partial^2 A}{\partial z^2} = 0 \textrm{,}  
\end{equation}
for which the corresponding boundary integral form can be written as
\begin{equation}
\label{biea} \beta\left(\bm{r'}\right) A(\bm{r'}) =  \text{P.V.}\oint_{\varGamma}[G_{0}(\bm{r'},\bm{r})\frac{\partial A}{\partial n}(\bm{r}) + A(\bm{r})\frac{\partial G_{0}}{\partial n}(\bm{r'},\bm{r})]dl(\bm{r}) \textrm{,}
\end{equation}
similar to equation \eqref{bie}, which is used as the boundary condition to solve \eqref{atransport}. The constant $\varsigma\left( t\right) $ is determined by integrating \eqref{laplacea} in the exterior and applying the Gauss-divergence theorem to obtain the following constraint for A on $\varGamma$
\begin{equation}
\label{ceqn} \oint_{\varGamma}\frac{\partial A}{\partial n} dl = 0 \textrm{.}
\end{equation}
The above equation implies that the net mean streamwise current is zero. Equations \eqref{adefinition}, \eqref{atransport}, \eqref{biea} and \eqref{ceqn} form the closure for the problem of computing the $x$-averaged in-plane components $\bar{b}_{y}$ and $\bar{b}_{z}$.

The Fourier coefficient components $\hat{b}_{x}$, $\hat{b}_{y}$ and $\hat{b}_{z}$ obtained for $k=0,1,2..$ $N_{x}/2-1$ are transformed back to the real space using an inverse FFT operation, which completes the computation of the secondary magnetic field evolution at a given time step. The $\bm{b}$ field obtained is used to compute the $\bm{j}$ field according to \eqref{amperelaw} and subsequently the Lorentz force term $\bm{j} \times \bm{b}_{t}$ in the momentum balance \eqref{navierstokes} for the computation of the velocity field at the next time step. The computational procedure described here is conducive for easy parallelization due to the fact that the numerical scheme is based on solution in the Fourier space. The computation of the Fourier coefficients in the $k$-space can be
 performed independently by assigning (virtually) individual processors to each of the $k$-modes. Our particular implementation of this numerical procedure for the solution of the induction equation with the integral boundary conditions was done through a FORTRAN code with hybrid MPI-OpenMP parallelization, starting with an existing quasistatic MHD code (DUCAT) \cite{Krasnov:2011}. 

\section{Verification and comparative study}
\subsection{Verification in the limiting case of low $R_{m}$}
An ideal verification of the implementation of the computation procedure described in the previous section would involve comparison of numerical results at $R_{m}\sim1$ and higher obtained from this procedure to those obtained using 
a full MHD numerical code that solves for the magnetic field on a grid covering an extended domain. However, since this is not possible, we limit our scope rather to verification of 
the computational procedure in the quasistatic limit. In this section, we present results for the case when the magnetic Reynolds number is low i.e. $R_{m}\ll1$ that aid as a verification of the implementation of
the numerical procedure. 
It is customary to describe magnetohydrodynamics at low $R_{m}$ with the quasistatic or
 inductionless approximation. In the most common formulation (which will be referred as QS formulation hereafter), the current density field $\bm{j}$ is described by the Ohm's 
law \eqref{jeqn} with the electric field $\bm{e}$ being expressed as the gradient of a scalar potential, $\bm{e}=-\nabla \phi$. The potential $\phi$ is obtained from the knowledge 
of the velocity field $\bm{v}$ and the divergence-free condition for the electric current density $\bm{j}$, by solving a Poisson equation. The QS formulation can be briefly 
summarized as
\begin{eqnarray}
\label{jeqn} \bm{j}&=& -\nabla\phi + \left(\bm{v} \times \bm{b}_{0} \right) \textrm{,} \\
\nabla^2\phi&=&\nabla \cdot \left(\bm{v} \times \bm{b}_{0} \right) \hskip3mm \text{Boundary condition:} \hskip2mm \frac{\partial \phi}{\partial n} = 0 \textrm{,} \\ 
\bm{f}_{L}&=& \frac{{\Har}^2}{\Rey}\left(\bm{j}\times\bm{b}_{0} \right) \textrm{,} 
\end{eqnarray}
where $\bm{f}_{L}$ is the Lorentz force source term in the Navier-Stokes equation \eqref{navierstokes} and the boundary condition corresponds to perfectly insulating walls. An alternative
formulation of the quasistatic approximation is the induced electric current based formulation that uses the current density $\bm{j}$ as the primary variable instead of the electric 
potential $\phi$ (see Smolentsev et al. \cite{Smolentsev:2010}).

\begin{figure}[!h]
        \subfigure[]{
                \centering
                \includegraphics[width=0.52\textwidth]{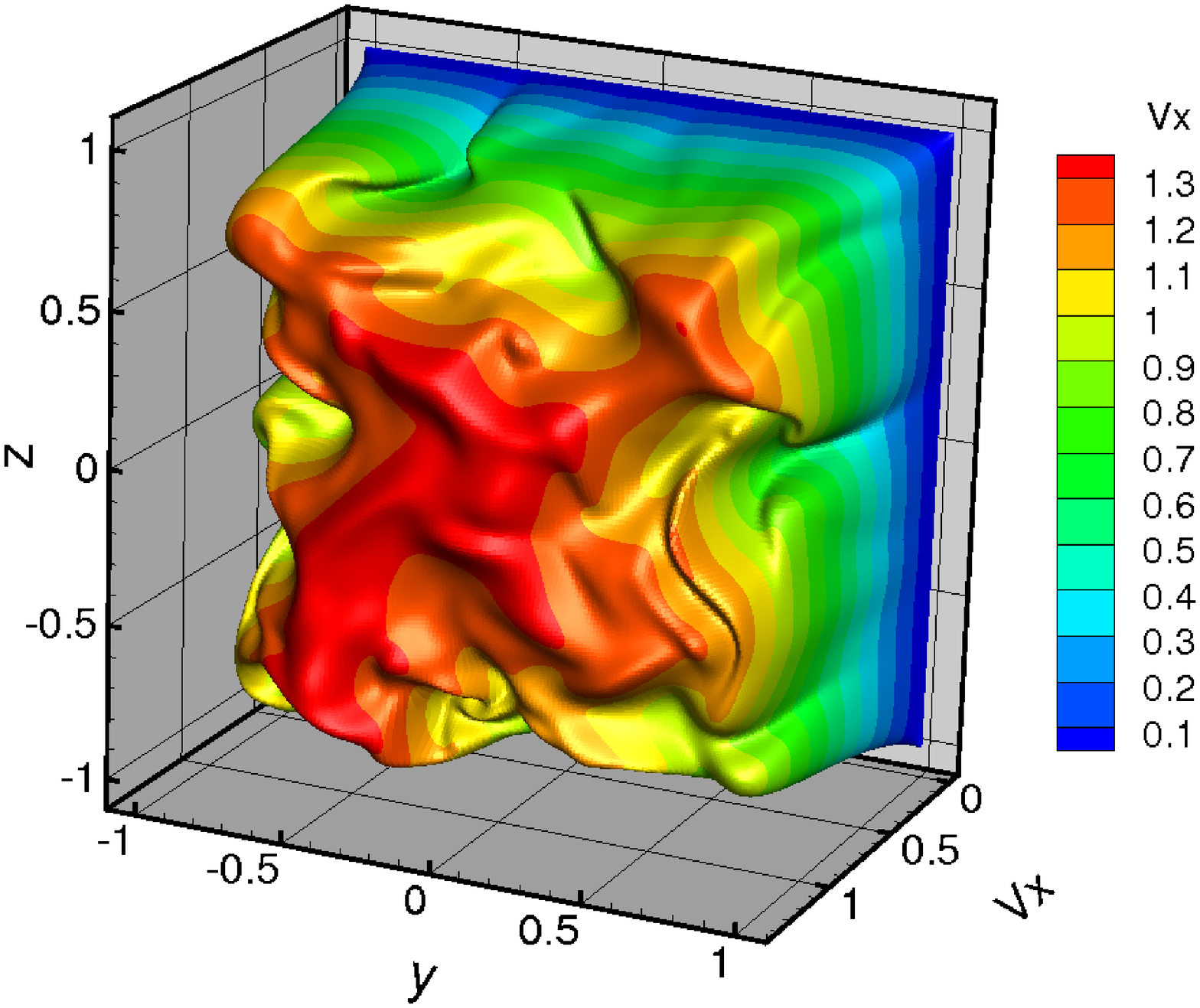}
        }
        \subfigure[]{
                \centering
                \includegraphics[scale=0.67]{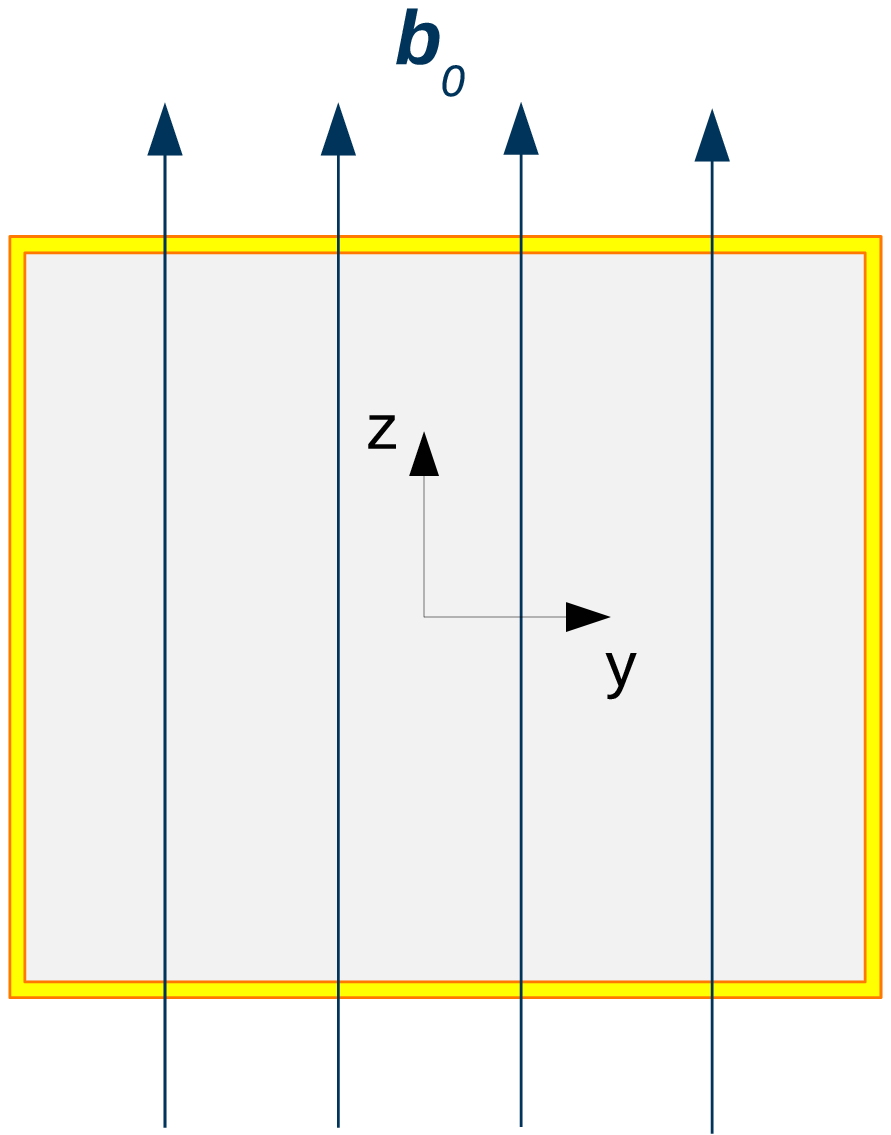}
        }

\caption{(a) Contour of the $x$-component of the initial turbulent velocity field at $\Rey=2000$ and (b) imposed magnetic field $\bm{b}_{0}=b_{0} \bm{k}$, shown at the cross-section $x=L_{x}/2$; $\Har=15$.}
\label{fig:initialfield}
\end{figure}

Furthermore, when $R_{m}$ is low, the secondary magnetic field is neverthless finite and its evolution can be described by another formulation of the quasistatic approximation
 based on the induced magnetic field rather than on the electric potential. This is the so called quasistationary formulation (referred to as QST formulation hereafter). The QST formulation can be obtained as follows. Approximating the electromagnetic fields by $\bm{\varpi}=\bm{\varpi}^0 + \varepsilon \bm{\varpi}^1$ where $\varepsilon=R_{m}$ is considered to be a small value and $\bm{\varpi}=\left[\bm{b},\bm{e},\bm{j} \right]$ denoting the magnetic, electric and the current density fields respectively, the induction equation can be rewritten as 
% Induction equation with expansion
\begin{equation}
\varepsilon \frac{\partial}{\partial t}\left(\bm{b}^0 + \varepsilon \bm{b}^1 \right) = \varepsilon \nabla \times \left( \bm{v}\times\left(\bm{b}^0 + \varepsilon \bm{b}^1  \right) \right) + \nabla^2 \left(\bm{b}^0 + \varepsilon \bm{b}^1 \right) \textrm{.}
\end{equation} 
Equating terms of the same order of $\varepsilon$ and assuming the imposed magnetic field $\bm{b}^0$ to be time-independent, we obtain
% Quasistationary equation
\begin{equation}
\label{quasistationary} \nabla^2\bm{b}^1  = (\bm{v}\cdot \nabla)\bm{b}^{0} - (\bm{b}^{0}\cdot \nabla)\bm{v} \textrm{,} \hskip5mm \bm{j}^1 = \nabla \times \bm{b}^1 \textrm{,}
\end{equation} 
 with the same integral wall boundary conditions for the magnetic field as described in the previous sections.  Through \eqref{quasistationary}, the magnetic field is parametrically dependent on time and evolves as a passive vector field that depends on the velocity field. It can be shown that $\nabla \times \bm{e}^1={\partial \bm{b}^0}/{\partial t}=0$, making the electric field expressible as $\bm{e}^1=-\nabla\phi$, through which the exact equivalence between the QS and QST formulations is established (see \cite{Boeck-Habilitation}) . Due to this equivalence, the current densities $\bm{j}$ computed by the QS formulation and the resulting secondary magnetic field must match with those computed by the QST formulation. 

% graphs showing jx, jy and jz comparison

\begin{figure}[!h]
        \subfigure[]{
          \includegraphics[width=0.5\textwidth]{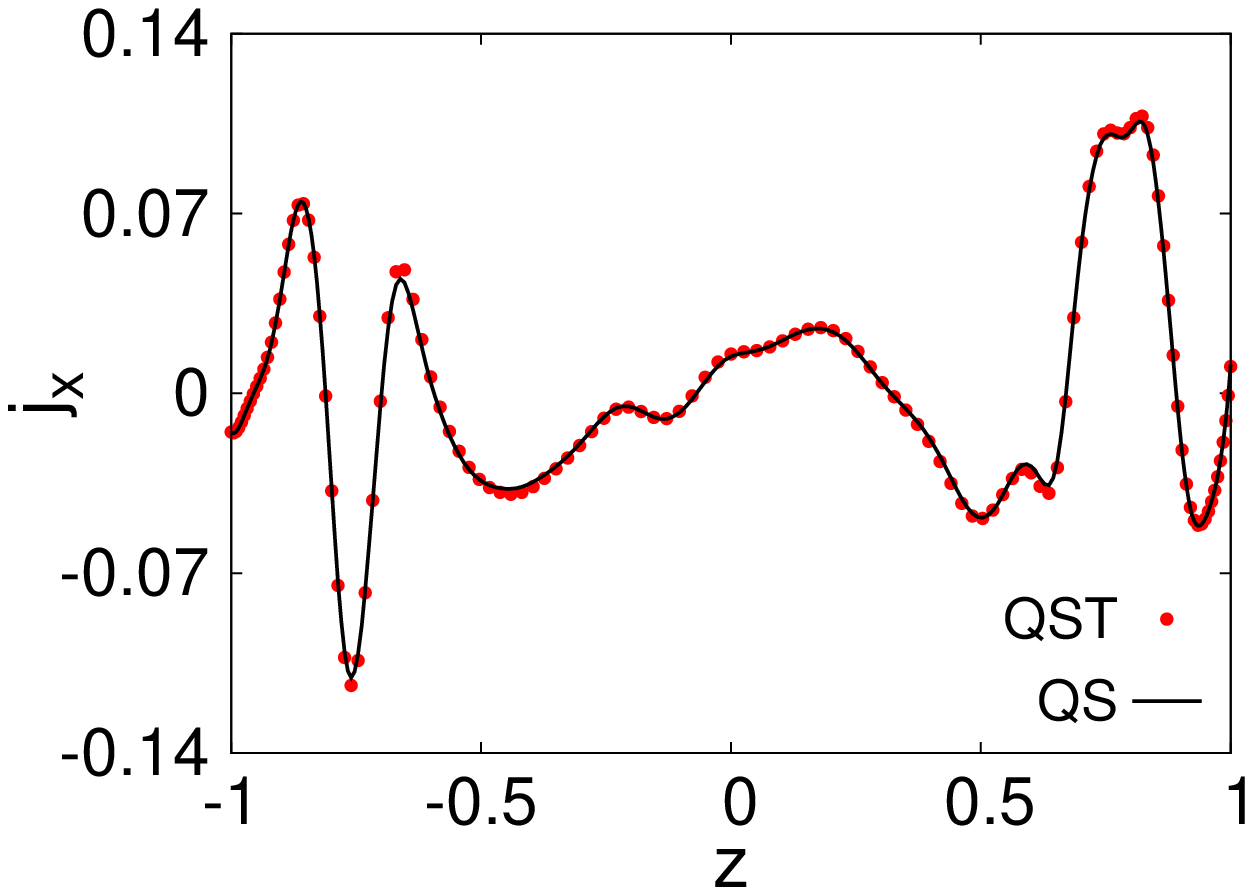}
          \label{figa}
        }
        \subfigure[]{
           \includegraphics[width=0.5\textwidth]{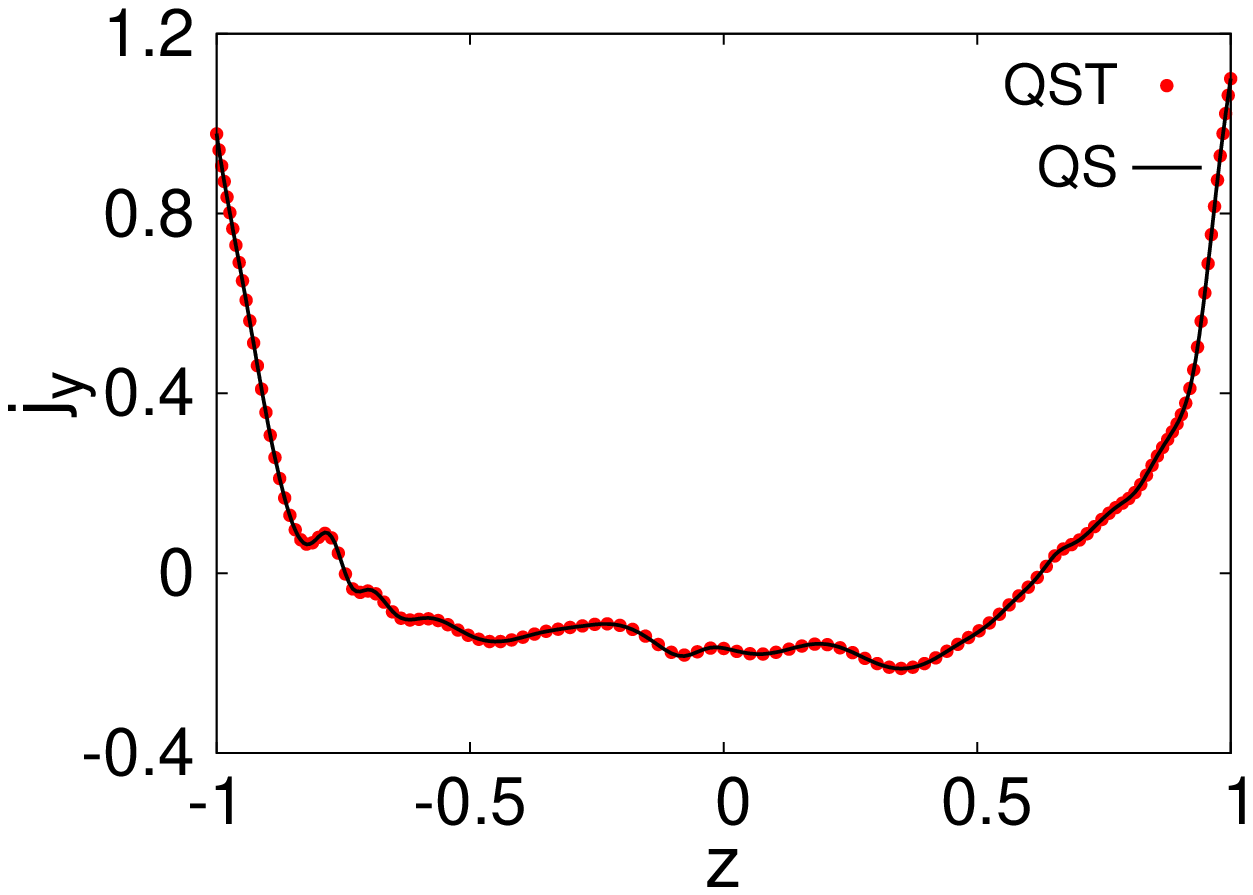}
           \label{figb}
        }

        \subfigure[]{
           \includegraphics[width=0.5\textwidth]{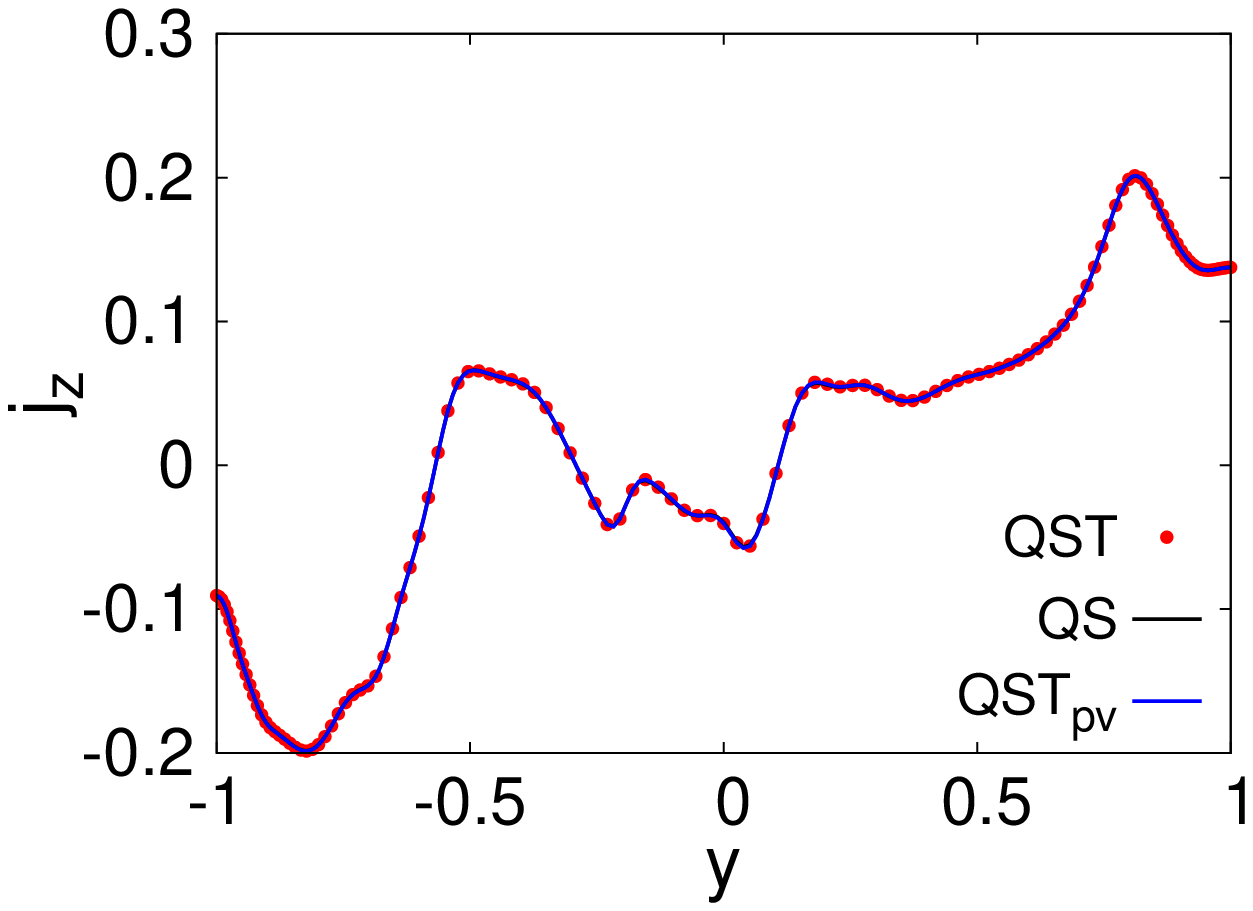}
           \label{figc}
        }
        \subfigure[]{
           \includegraphics[width=0.5\textwidth]{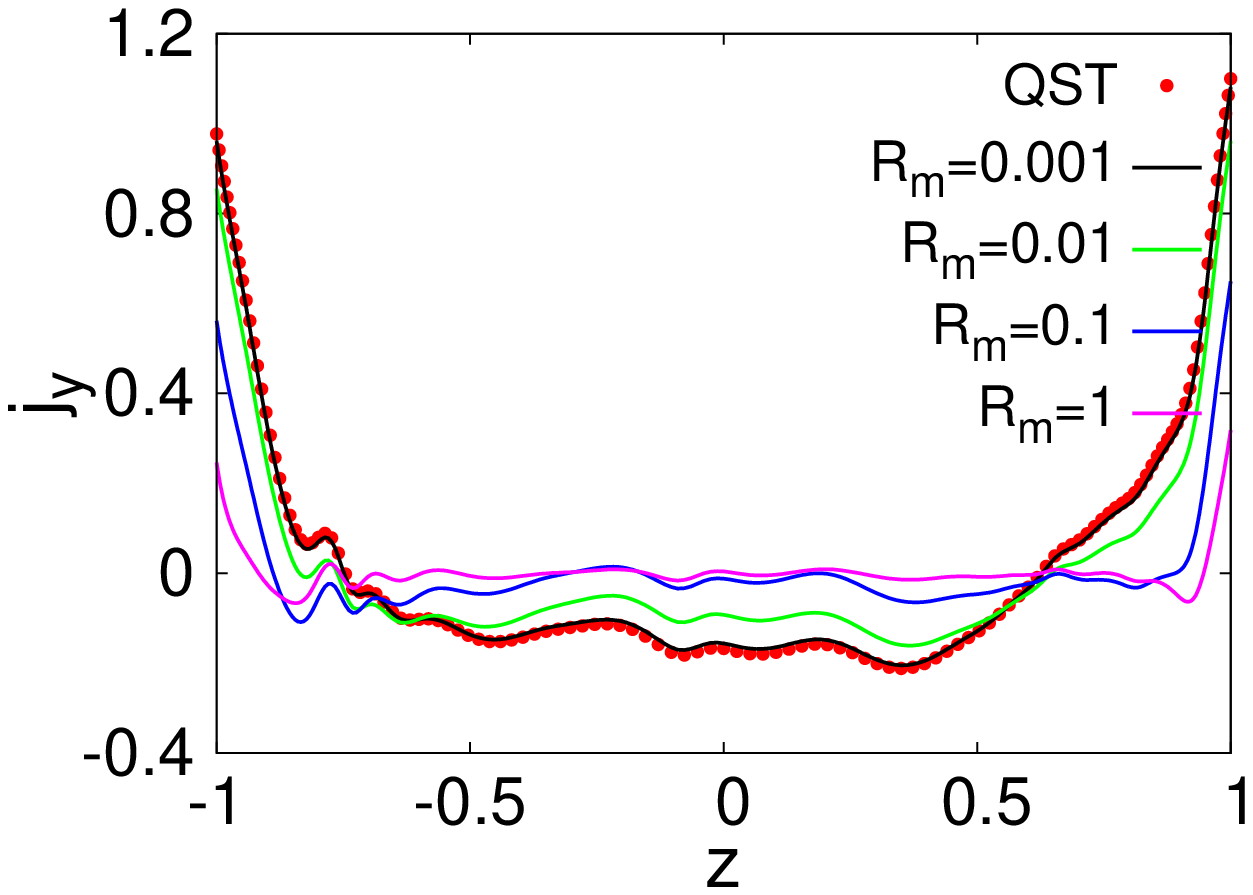}
           \label{figd}
        }
\caption{Current density components plotted at the cross-section $x=L_{x}/2$. (a) $j_{x}$ and (b) $j_{y}$ along the line $y=0$; (c) $j_{z}$  along the line $z=-0.5$; (d) $j_{y}$ from full MHD. Grid: $256\times256\times256$, $\Rey=2000$, $\Har=15$.}
\label{fig:jxjyjz}
\end{figure}

In the particular case that we consider, a uniform magnetic field along the $z$-direction is imposed on a fully turbulent 3D velocity field at $ Re= 2000 $ (see Fig.~\ref{fig:initialfield}) in a duct of length $L_{x}=4\pi$ and a square cross-section $L_{y}=L_{z}=2$ and the numerical computation is performed for a single time step with both the QS and the QST procedures. A grid resolution of $256^3$ is used for this computation. In order to perform the inductionless computations, the quasistatic MHD code DUCAT (DNS code based on finite differences) was used which has been extensively validated (see \cite{Krasnov:2011}). The resulting components of current densities from the two methods are compared at a particular cross section ($x=L_{x}/2$) as shown in Fig.~\ref{figa} and \subref{figb} and a close match between the two methods is observed. However, it must be mentioned that a good agreement of current densities is only a neccesary requirement for the correctness of the quasistationary procedure with BEM but not a sufficient one. This is attributed to the fact that in the case of low magnetic Reynolds number, when $j_{n}=0$ is ensured on the wall, the current density field $\bm{j}$ in the interior is uniquely determined. Due to this reason, the current densities will match even if a simplified approach, the so called pseudo-vacuum magnetic boundary conditions (explained in the next subsection), are applied to the quasistationary formulation. This is shown in Fig.~\ref{figc}, where the component $j_{z}$ shows a good agreement between the QS, QST and the QST with pseudo-vacuum BCs which is denoted as QST$_{\text{pv}}$ in the legend. A proof for the uniqueness of $\bm{j}$ in the case of $j_{n}=0$ is provided in the Appendix.

Of particular interest is the order of $R_m$ at which the validity of the quasistatic approximation really hold. For this purpose, the full MHD system (the induction equation) with the integral boundary conditions was used to compute the $\bm{b}$ field for a single time step at various orders of $R_m$. The resulting current component $j_{y}$ obtained is compared with that obtained from the QST formulation. It can be seen (from Fig.~\ref{figd}) that a convergence to the quasistatic limit occurs when $R_m\sim10^{-3}$.  

\begin{figure}[!h]
        \subfigure[]{
          \includegraphics[width=0.5\textwidth]{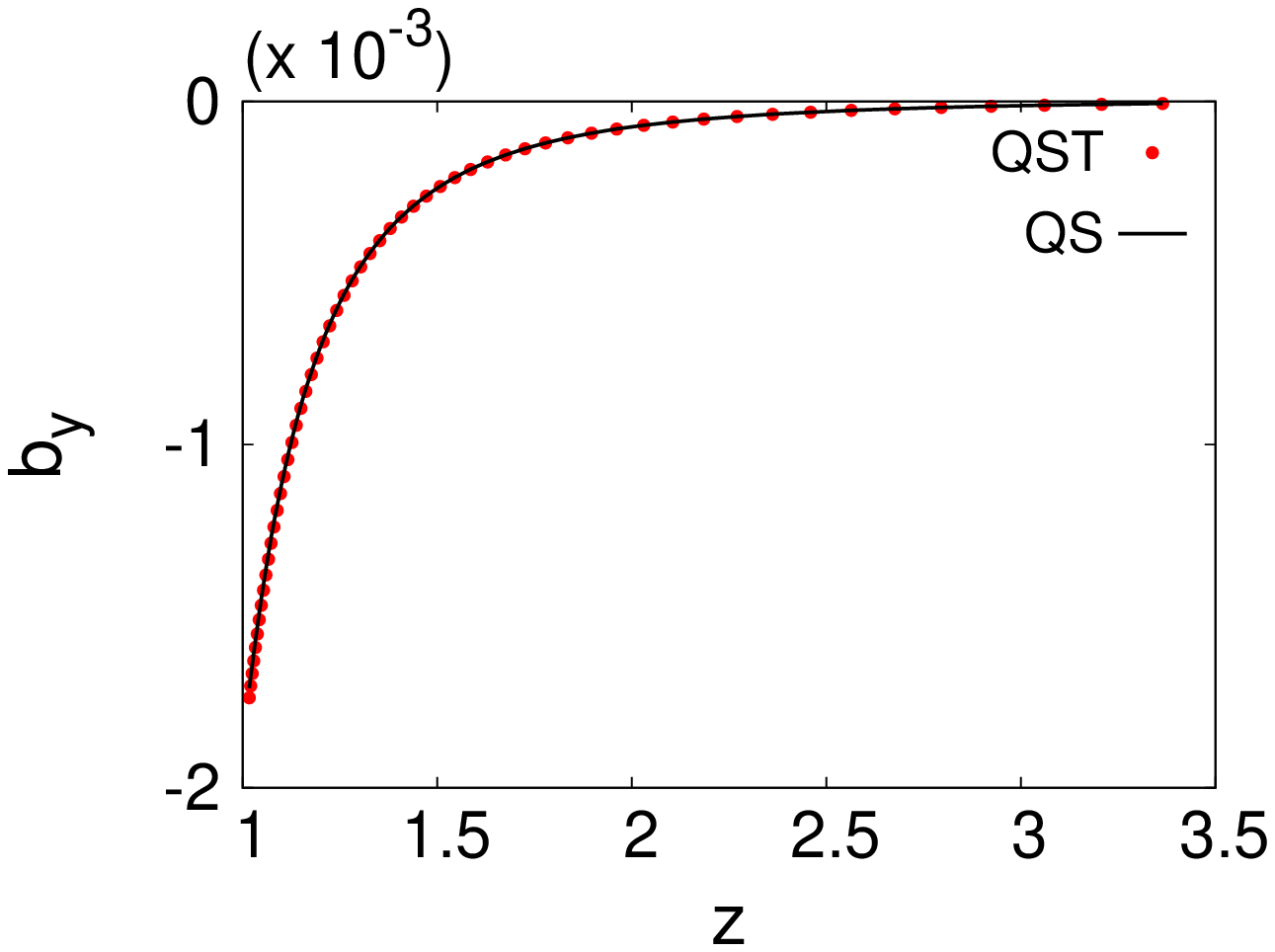}
          \label{fig:by_ext}
        }
        \subfigure[]{
          \includegraphics[width=0.5\textwidth]{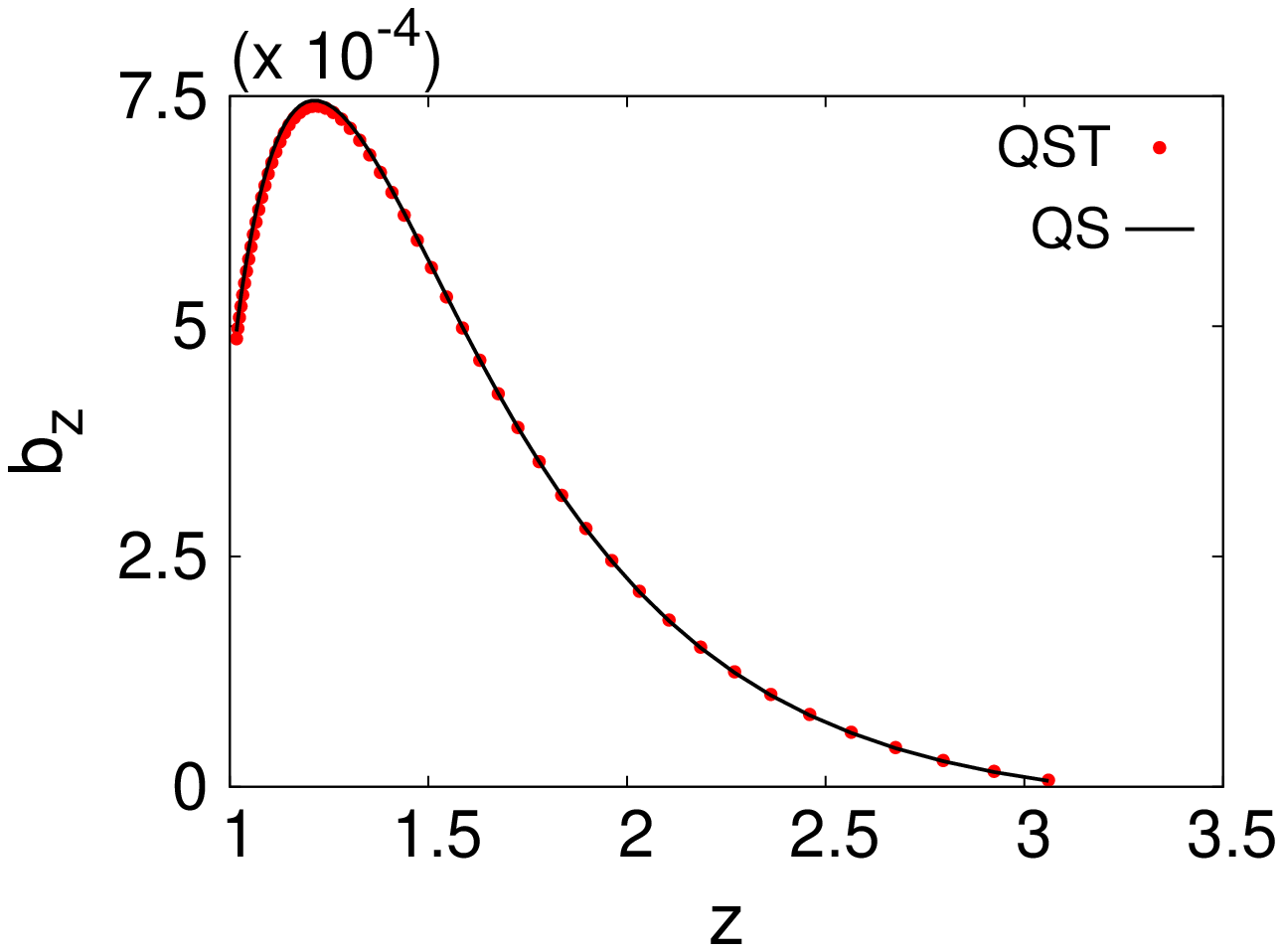}
          \label{fig:bz_ext}
        }

        \subfigure[]{
          \includegraphics[width=0.5\textwidth]{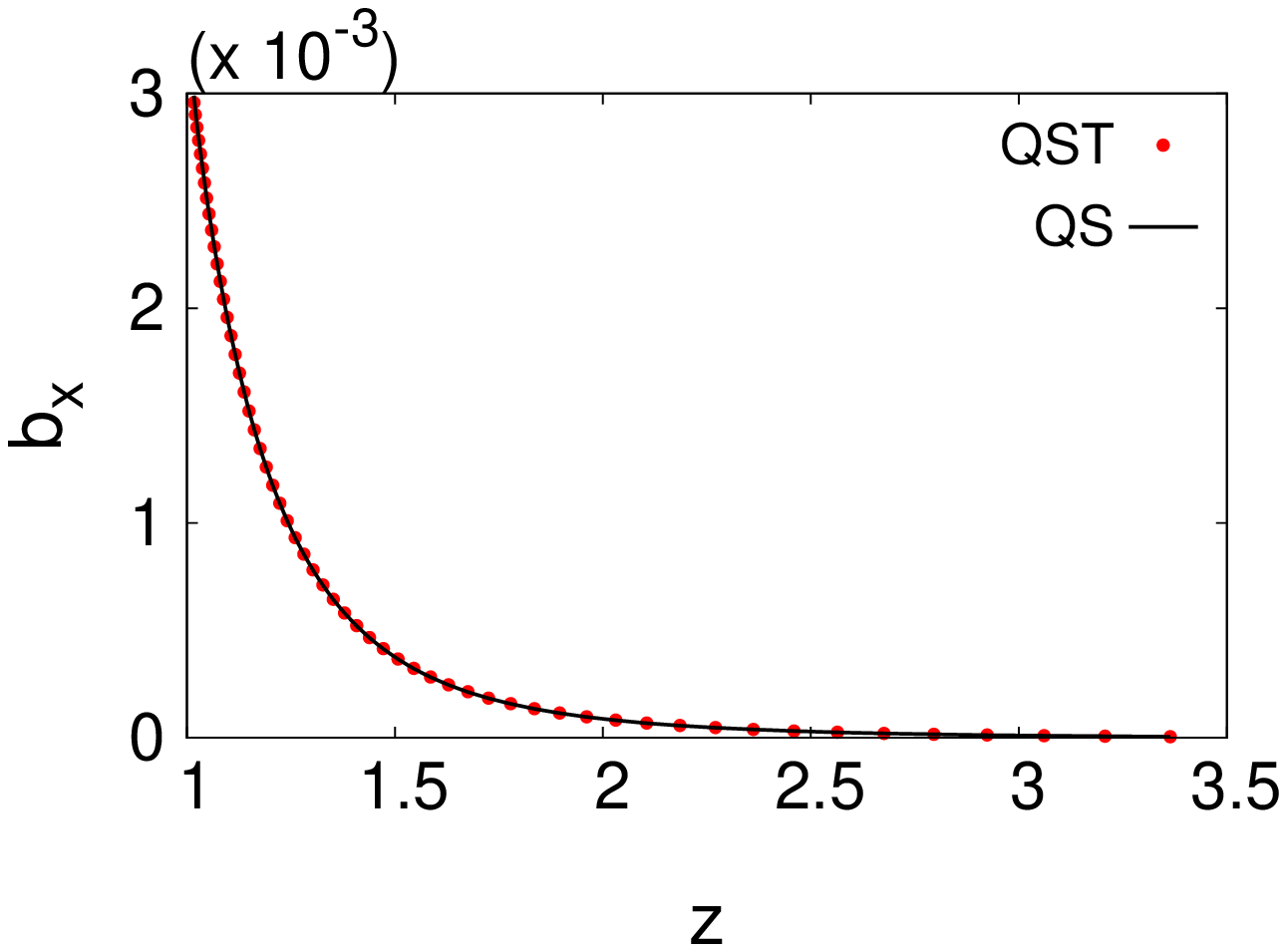}
          \label{fig:bx_ext}
        }
        \subfigure[]{
          \includegraphics[width=0.5\textwidth]{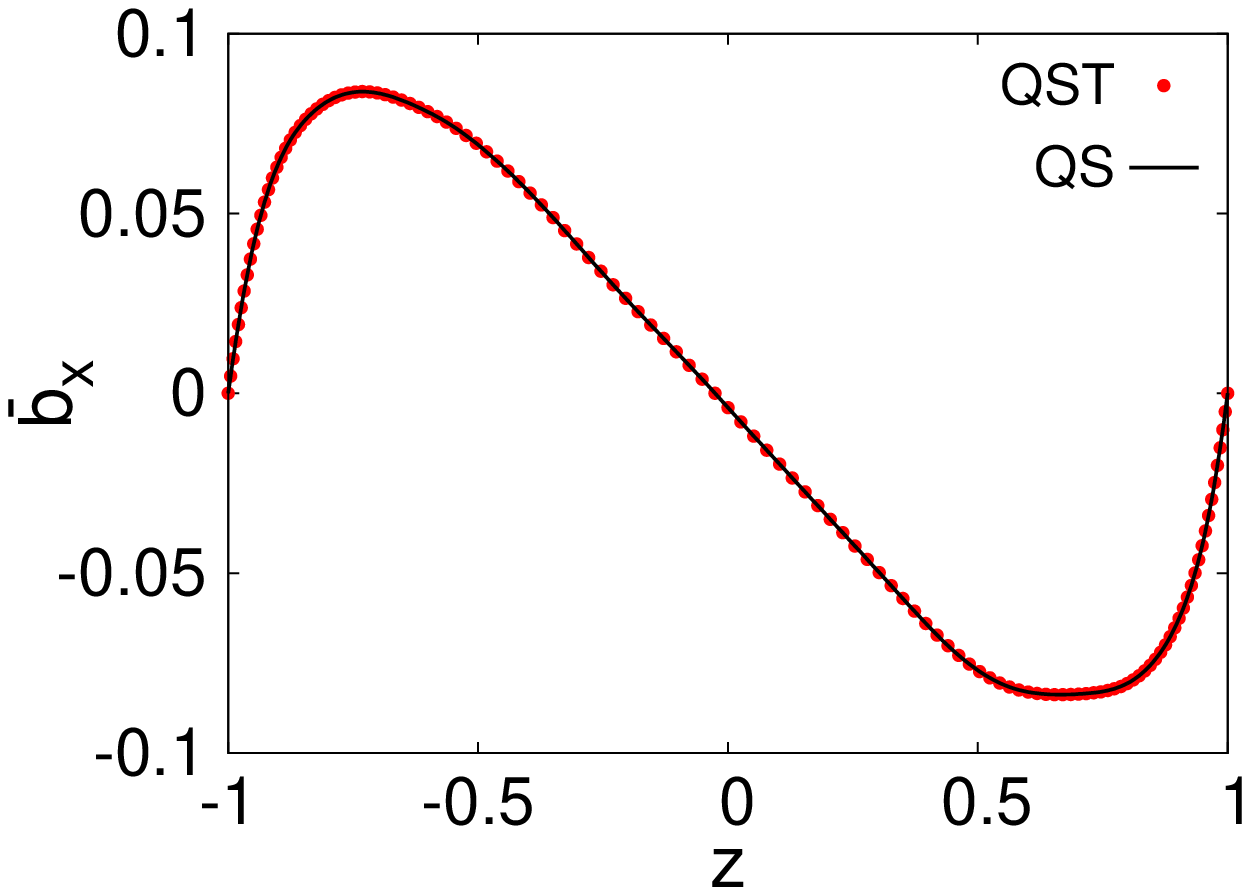}
          \label{fig:bxm_ext}
        }

\caption{Secondary magnetic field components in the exterior (a) $b_{y}$ (b) $b_{z}$ and (c) non-zero modes of $b_{x}\left(k\neq0\right)$, at the streamwise location $x=L_{x}/2$. The exterior corresponds to $z>1$. (d) Mean streamwise component of the secondary magnetic field $b_{x}\left(k=0\right)$ in the interior of the duct; Grid: $256\times256\times256$, $\Rey=2000$, $\Har=15$.}

\label{fig:bxbybz_ext}
\end{figure}

 To complement the verification, a comparison is made for the secondary magnetic field in the exterior of the duct. For this, the current density field $\bm{j}$ obtained from the quasistatic computation is used to compute the secondary magnetic field in the duct exterior through the Biot-Savart law
% Biot-Savart law
\begin{equation}
\bm{b}\left(\bm{r'}\right) =\frac{1}{4\pi}\int\frac{\bm{j}\left(\bm{r}\right) \times\left( \bm{r'}-\bm{r}\right) dV}{{\left( \lvert \bm{r'}-\bm{r}\rvert\right)}^3} \textrm{,}
\end{equation}
which is evaluated numerically using a trapezoidal quadrature. The corresponding magnetic field from the quasistationary computation is obtained by evaluating the scalar potential $\hat{\psi}_{k}$ in the duct exterior using  equation \eqref{bie} from the known values of $\hat{\psi}_{k}$ and $\hat{b}_{nk}$ on the boundary but with $\beta\left(\bm{r'}\right)=1$. A comparison of the exterior field components $b_{y}$ and $b_{z}$ along the line $y=0,~z>1$ is shown in Fig.~\ref{fig:by_ext} and \subref{fig:bz_ext} respectively.

The streamwise component $b_{x}$ is decomposed into $b_{x}\left(k\neq0\right)$ and $b_{x}\left(k=0\right)$ that contain the non-zero modes and the zero mode respectively for which the comparison is shown in Fig.~\ref{fig:bx_ext} and \subref{fig:bxm_ext}. Since the mean component $\bar{b}_{x}$ vanishes in the exterior, its comparison is made only in the interior of the duct. This concludes the verification of the numerical procedure adopted to model the magnetic boundary conditions for the induction equation.

\subsection{Comparison with pseudo-vacuum boundary conditions}
As mentioned in section 1, finite/high $R_{m}$ MHD simulations are often conducted using the so called pseudo-vacuum magnetic boundary conditions (\cite{Hurlburt:1988,Brandenburg:1995,Rüdiger:2001,Gailitis:2004,Kenjeres:2007,Hubbard:2010}) 
\begin{equation}
\textrm{Pseudo-vacuum BCs :} \hskip5mm \bm{b}_{\parallel} = 0,\: \frac{\partial b_{n}}{\partial n} = 0 \hskip2mm\mbox{at } y,z=\pm 1 \textrm{,}
\end{equation}
where the subscript $\parallel$ refers to the two wall tangential directions. 
\begin{figure}[!h]
\centering
        \subfigure[]{
                \includegraphics[width=0.31\textwidth]{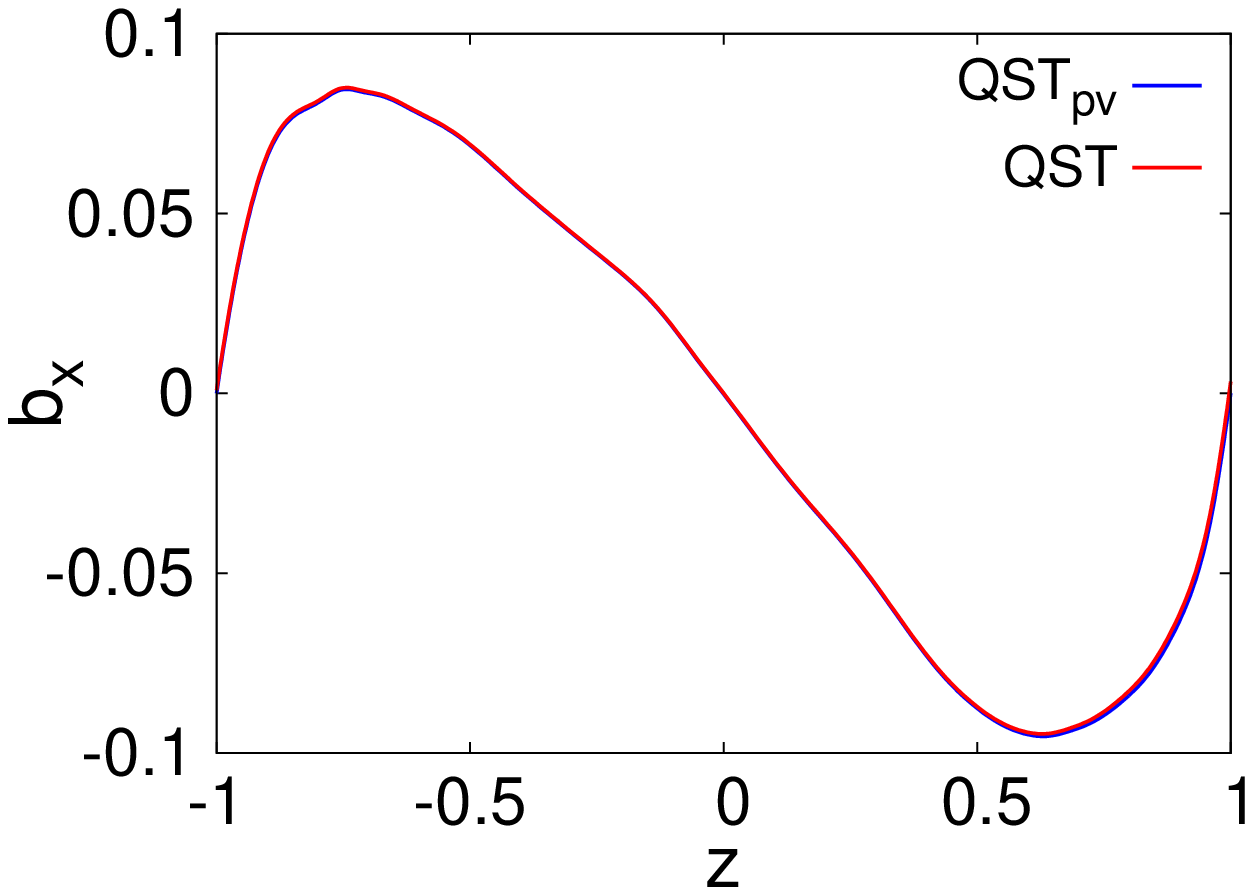}
        }
        \subfigure[]{
                \includegraphics[width=0.31\textwidth]{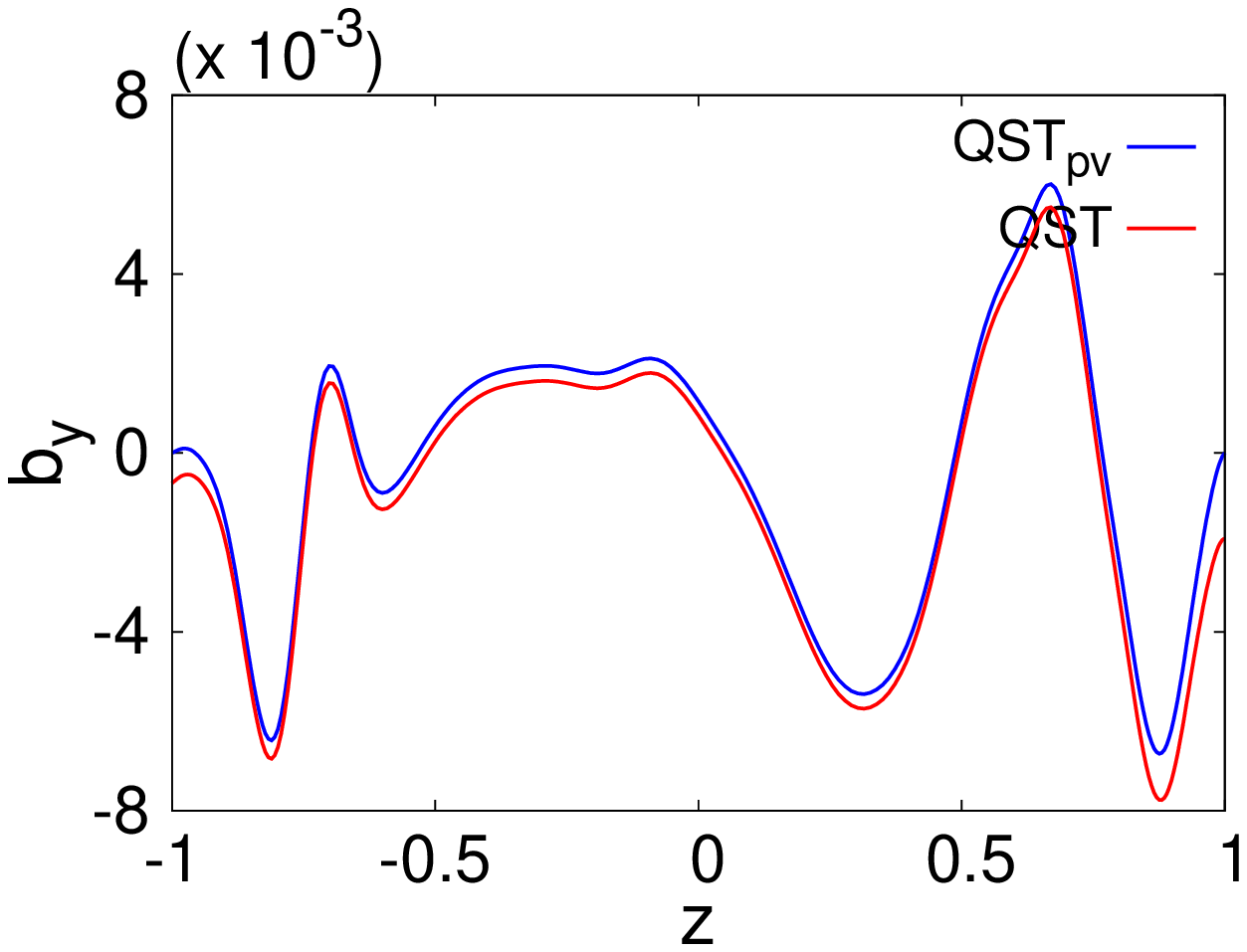}
        }
        \subfigure[]{
                \includegraphics[width=0.31\textwidth]{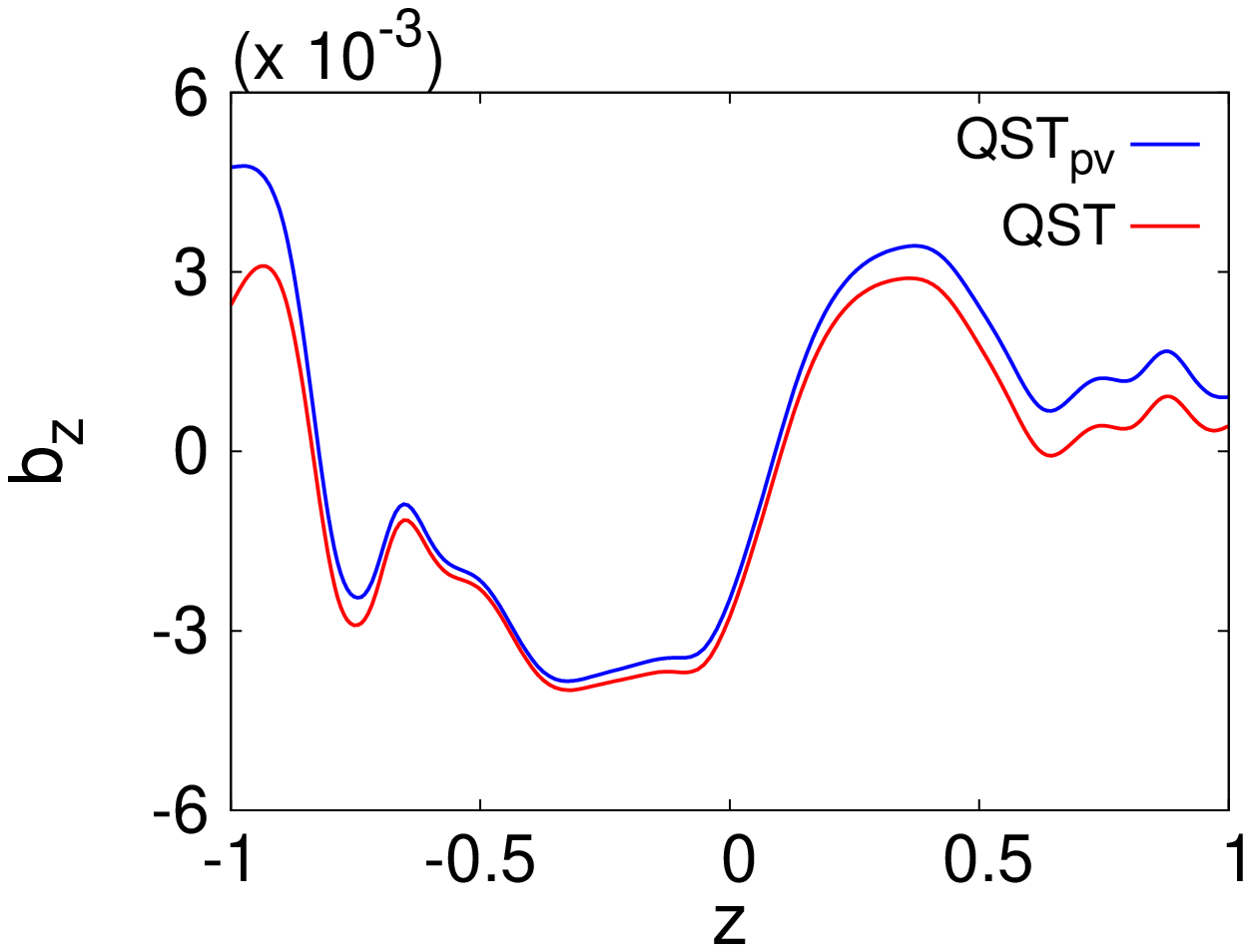}
        }
\caption{Secondary magnetic field components (a) $b_{x}$ (b) $b_{y}$ and (c) $b_{z}$ along the line $y=0$ on the cross-sectional plane $x=L_{x}/2$; Grid: $256\times256\times256$, $\Rey=2000$, $\Har=15$.}
\label{fig:bxbybz_pv}
\end{figure}
This formulation achieves vanishing wall normal currents $j_{n}$ through the assumption of zero tangential magnetic field, a trivial solution of $\left( \nabla\times\bm{b}_{wall}\right)\cdot\bm{n}=0$, and leads to considerable simplification of the computational procedure. However, numerical solutions obtained with this simpified model can result in significant loss of accuracy in the near wall velocity and magnetic fields. This becomes particularly important for wall-bounded MHD flows at transitional regimes, since instabilities are triggered in the thin boundary layers (either Shercliff layers that appear near the walls parallel to the magnetic field or Hartmann layers that appear near the walls perpendicular to the magnetic field). Here, differences that arise using the pseudo-vacuum conditions are quantified for the case of low $R_{m}$. In Fig.~\ref{fig:bxbybz_pv} magnetic field components in the duct interior computed using the boundary integral procedure are compared to those computed using the pseudo-vacuum conditions.

It is observed that the primary streamwise magnetic field component $b_{x}$ matches very well but the secondary components $b_{y}$ and $b_{z}$ show significant differences (especially near 
the walls) in both the cases. In this particular case of low $R_{m}$, the Lorentz force being proportional to $\bm{j}\times\bm{b}_{0}$, these differences do not impact the flow field.
 However, at finite/high $R_{m}$, the effect of these differences on the velocity field would likely be significant. 

\section{Turbulence evolution in a Hartmann duct flow}
In this section, the computational procedure outlined in this paper is applied to simulate and study certain specific features of turbulent duct flow in the presence of a uniform wall 
normal magnetic field 
at moderate magnetic Reynolds numbers $R_{m}=50$ and $R_{m}=100$ and at a low hydrodynamic Reynolds number $\Rey=5000$. We study 
the effect of $R_{m}$ on the evolution of turbulence
at relatively low Hartmann number and also on the relaminarisation of the flow at Hartmann numbers close to the critical/threshhold values. The aim of this study is to obtain a 
sense of the impact of $R_{m}$ on turbulent Hartmann duct flow. A comprehensive study of the dependencies on $R_{m}$ in larger Reynolds number and Stuart number flows will be given elsewhere. 
For this purpose, 
a purely hydrodynamic turbulent 
duct flow in a domain size of $4\pi\times2\times2$, that has evolved to a statistically steady state is chosen as the initial state and a uniform magnetic field along the $z$-direction, 
$\bm{b}_{0}=b_{0}\hat{k}$ is imposed on the flow. 
The subsequent evolution of the velocity and magnetic fields are computed on a grid size of $256\times192\times192$, with an equal grid stretch factor in the $y$ and $z$ directions,
$S_{y}=S_{z}=1.8$. It must be noted that, in practice when a magnetic field (generated by external current sources) is applied onto a conducting flow at finite magnetic Reynolds number, 
the field diffuses at a rate proportional to $\sqrt{\lambda}$, unlike the case of a low $R_{m}$ flow where the magnetic field diffuses instantly (relative to the time scales 
relevant to this problem) throughout the conducting medium. However, in order to have an initial state that allows direct comparison with the low $R_{m}$ case, we assume here that an intial
uniform magnetic field is present throughout even in the case of flows with moderate $R_{m}$.

\subsection{Relaminarization threshold}
Transition from a laminar flow to a turbulent flow or vice versa and the critical parameters at which this happens in Hartmann duct and channel flows has been of significant interest right from the time Hartmann performed 
his first experimental studies in 1937 (\cite{Hartmann:1937}). One of the reasons for this is the significant impact that transition to turbulence can have on quantities of engineering interest like the skin fricion factor.
Since then, numerous experiments
and in the recent decades, several numerical studies have been conducted that lead to a better understanding of transition in Hartmann 
duct and channel flows at low magnetic Reynolds 
numbers (see for e.g. \cite{Murgatroyd:1953,Reed:1978,Kobayashi:2008,Dmitry:2013}). However, the effect of finite magnetic Reynolds 
number on the suppression of duct flow turbulence by a magnetic field is unknown, which we explore here. To this end,
we simulate the evolution of a turbulent duct flow at $\Rey=5000$ in the presence 
\begin{figure}[!h]
   \includegraphics[width=1.0\textwidth]{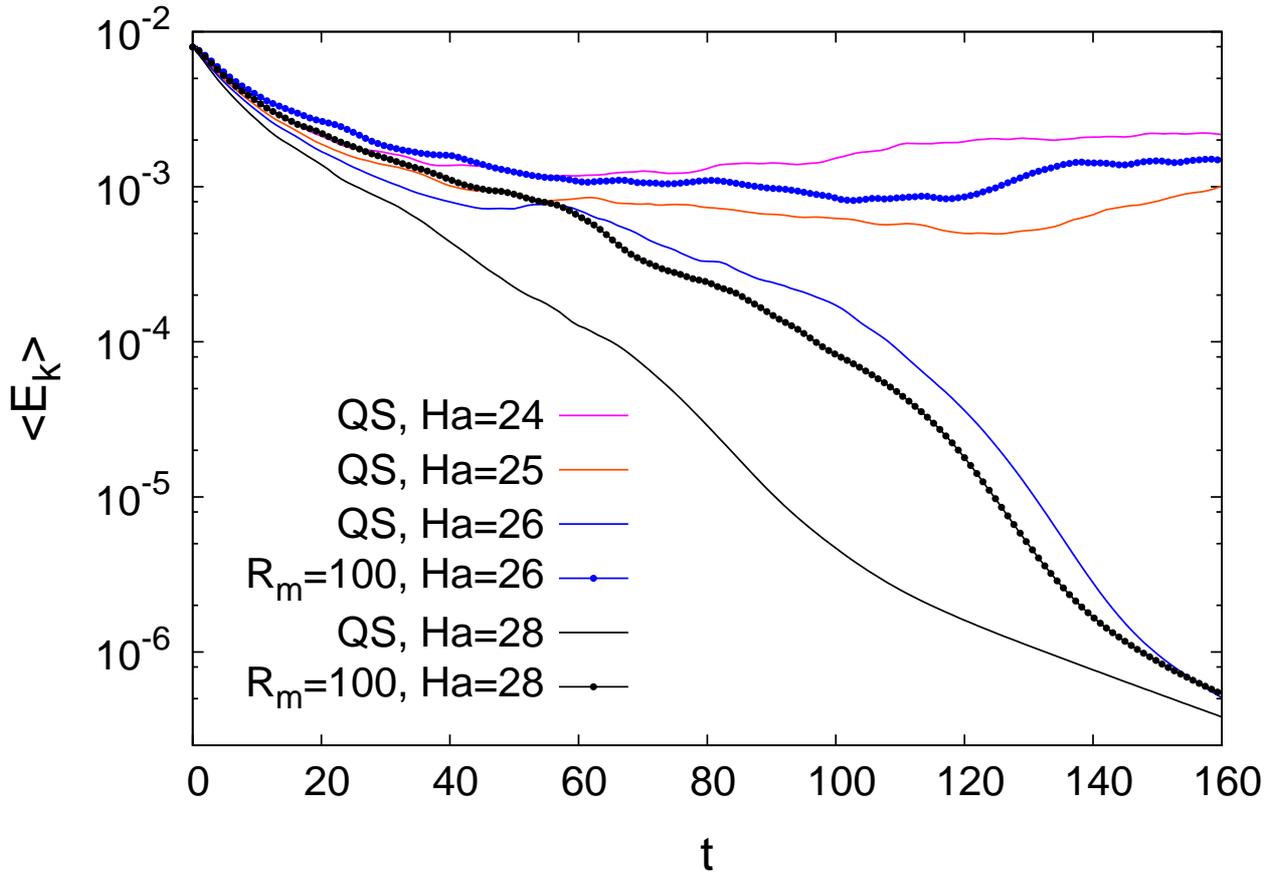}
\caption{Evolution of turbulent kinetic energy $\langle E_{k}\rangle$ at $\Rey=5000$ and Hartmann numbers close to the relaminarisation threshold.}
\label{fig:ek_relam}
\end{figure}
of a uniform intial magnetic field at $R_{m}=100$ and at Hartmann numbers close to the critical values ($Ha\approx25$) 
obtained from quasistatic ($R_{m}\ll1$) DNS studies performed by Krasnov et. al. (\cite{Dmitry:2013}).
It is observed from the QS simulations that the flow becomes laminar between $Ha=25$ and $Ha=26$, which can be clearly seen in Fig.~\ref{fig:ek_relam} from the near exponential decay 
of turbulent kinetic energy $\langle E_{k} \rangle$ to negligible values when $Ha=26$. Turbulent kinetic energy is defined in this case as,
\begin{equation}
\label{eq:tkenergy} \langle E_{k}\rangle=\frac{1}{8l_{x}}\int \limits_{0}^{l_{x}}\int \limits_{-1}^{1}\int \limits_{-1}^{1}\left({v'_{x}}^2+{v'_{y}}^2+{v'_{z}}^2\right)dxdydz,
\end{equation}
\begin{figure}[]
        \begin{center}
        \subfigure[]{
                \includegraphics[width=0.38\textwidth,trim=15mm 20mm 65mm 50mm,clip]{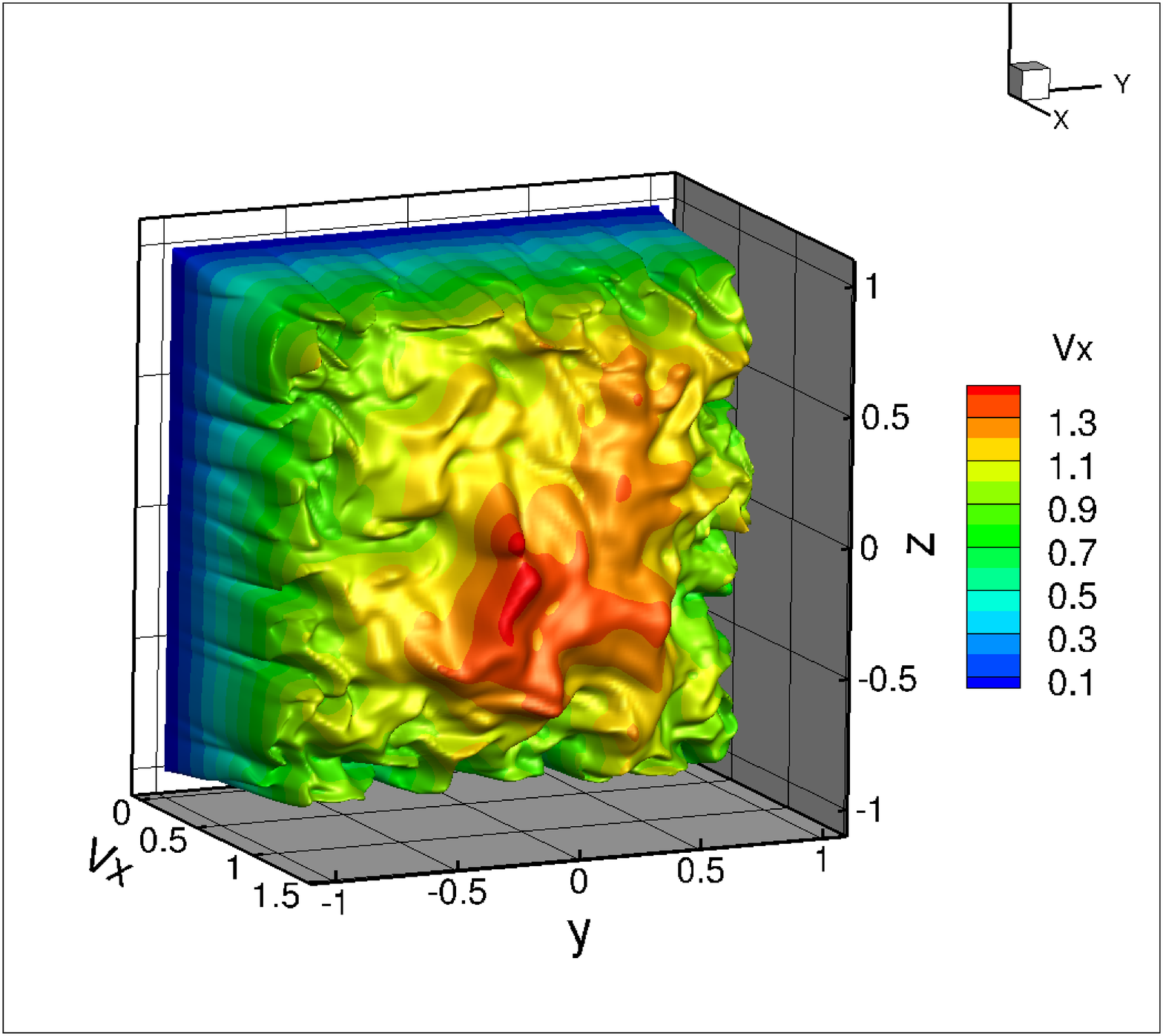}
        }
        \end{center} 
        \subfigure[]{
                \includegraphics[width=0.38\textwidth,trim=15mm 20mm 65mm 50mm,clip]{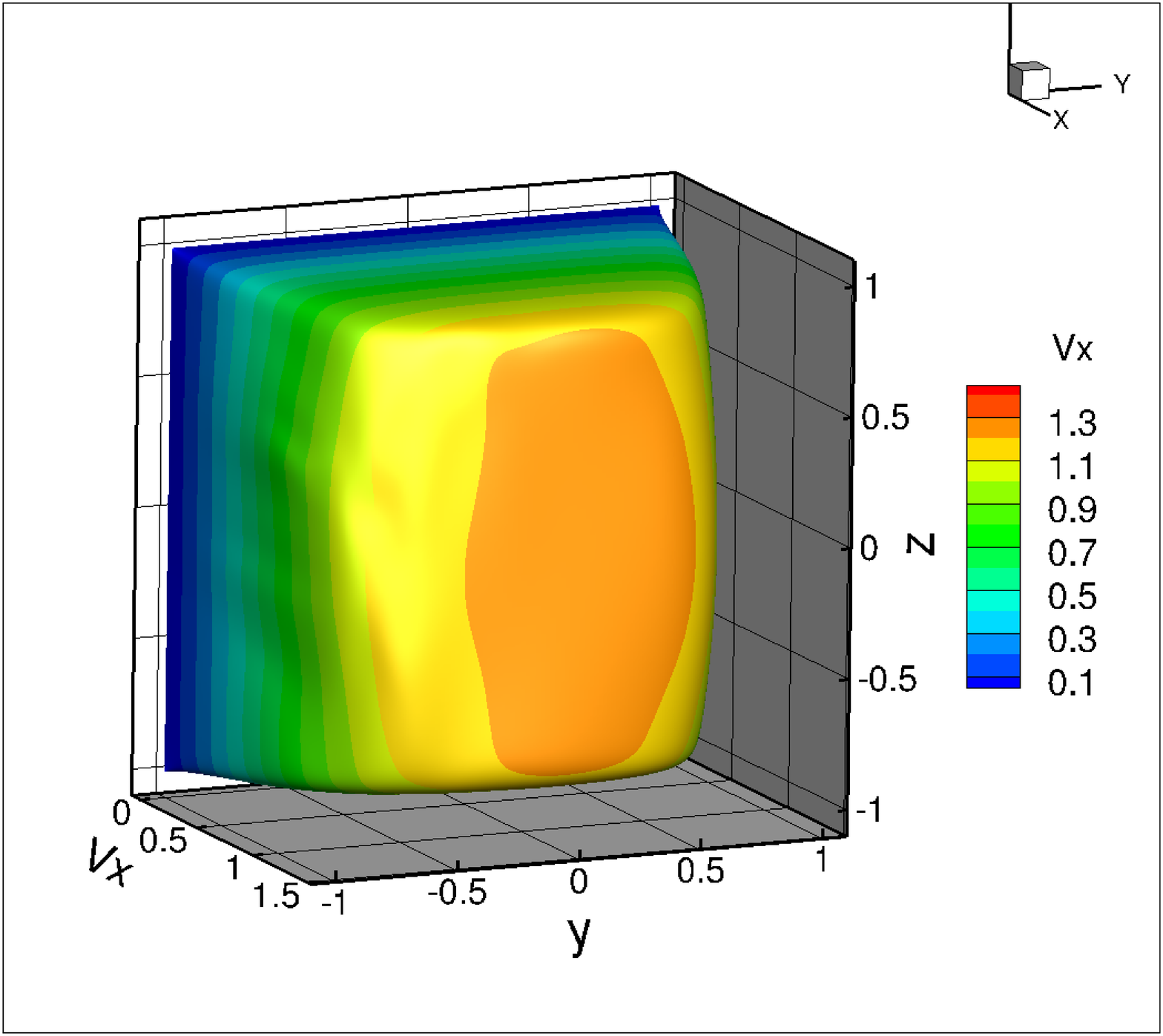}
                 \label{fig:rm0_ha26}
        }
        \subfigure[]{
                \includegraphics[width=0.38\textwidth,trim=15mm 20mm 65mm 50mm,clip]{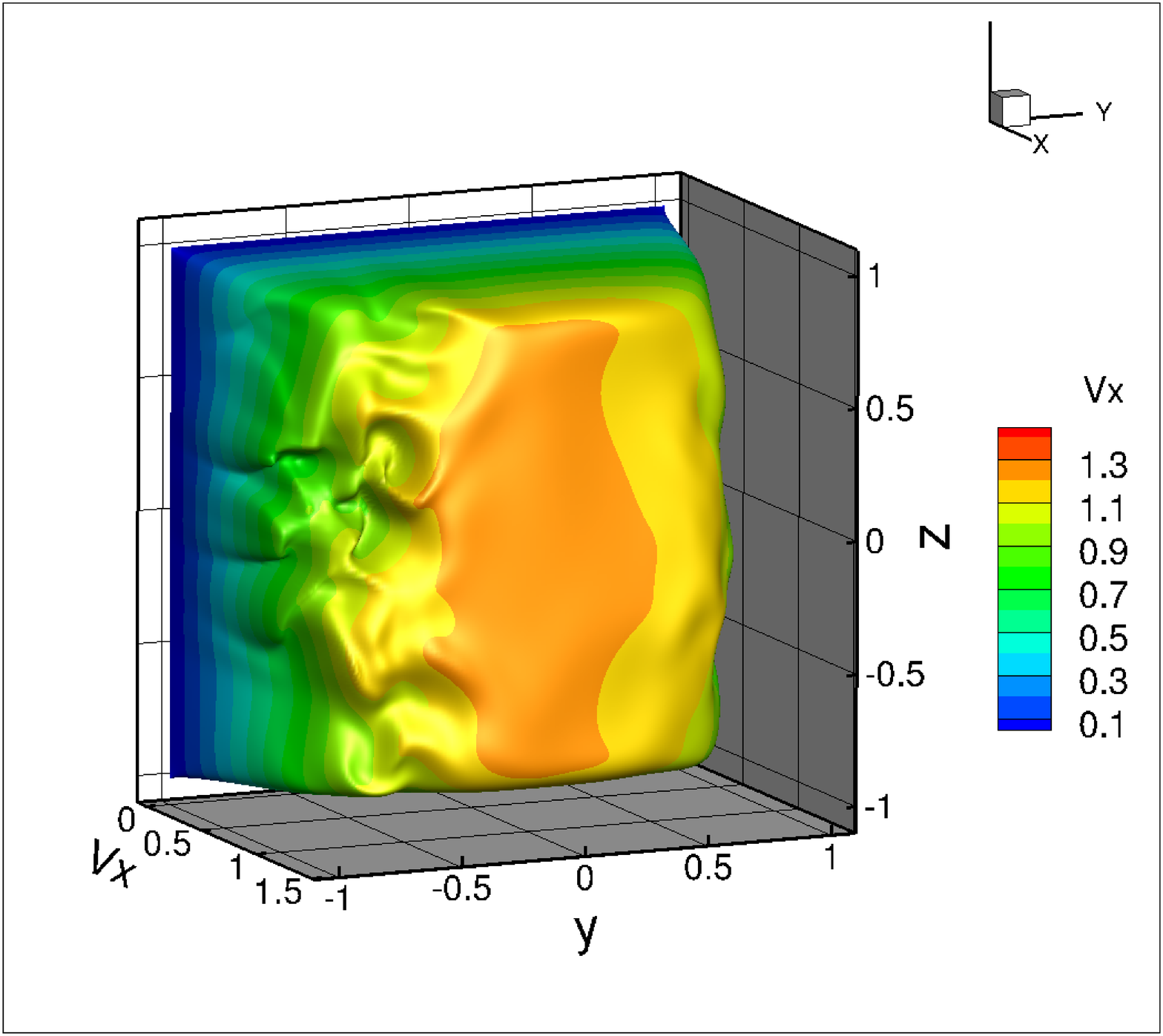}
                 \label{fig:rm100_ha26}
        }
        \subfigure{
                \includegraphics[width=0.12\textwidth,trim=320mm 100mm 5mm 112mm,clip]{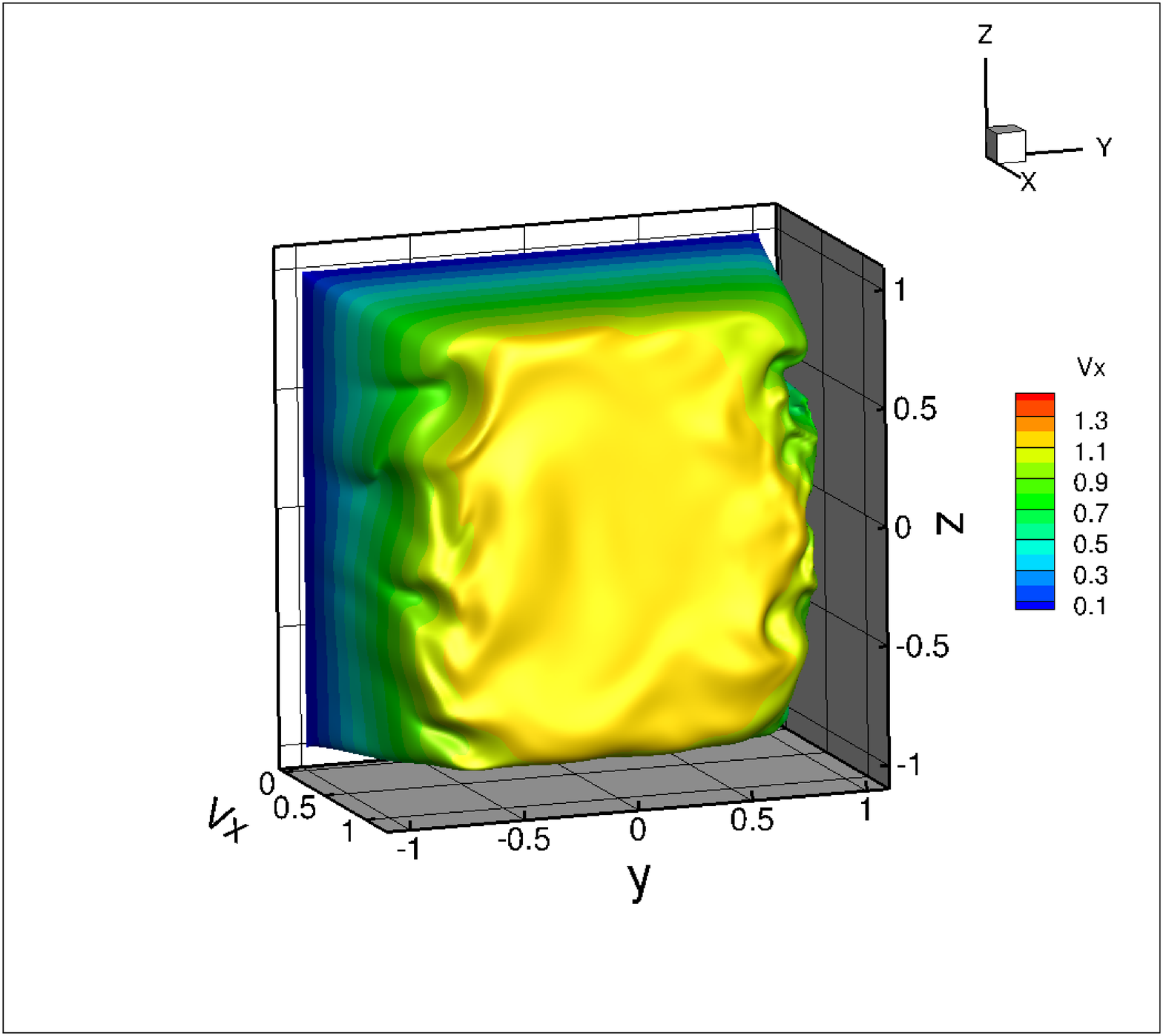}
        }
        \subfigure[]{
                \includegraphics[width=0.38\textwidth,trim=64mm 20mm 50mm 62mm,clip]{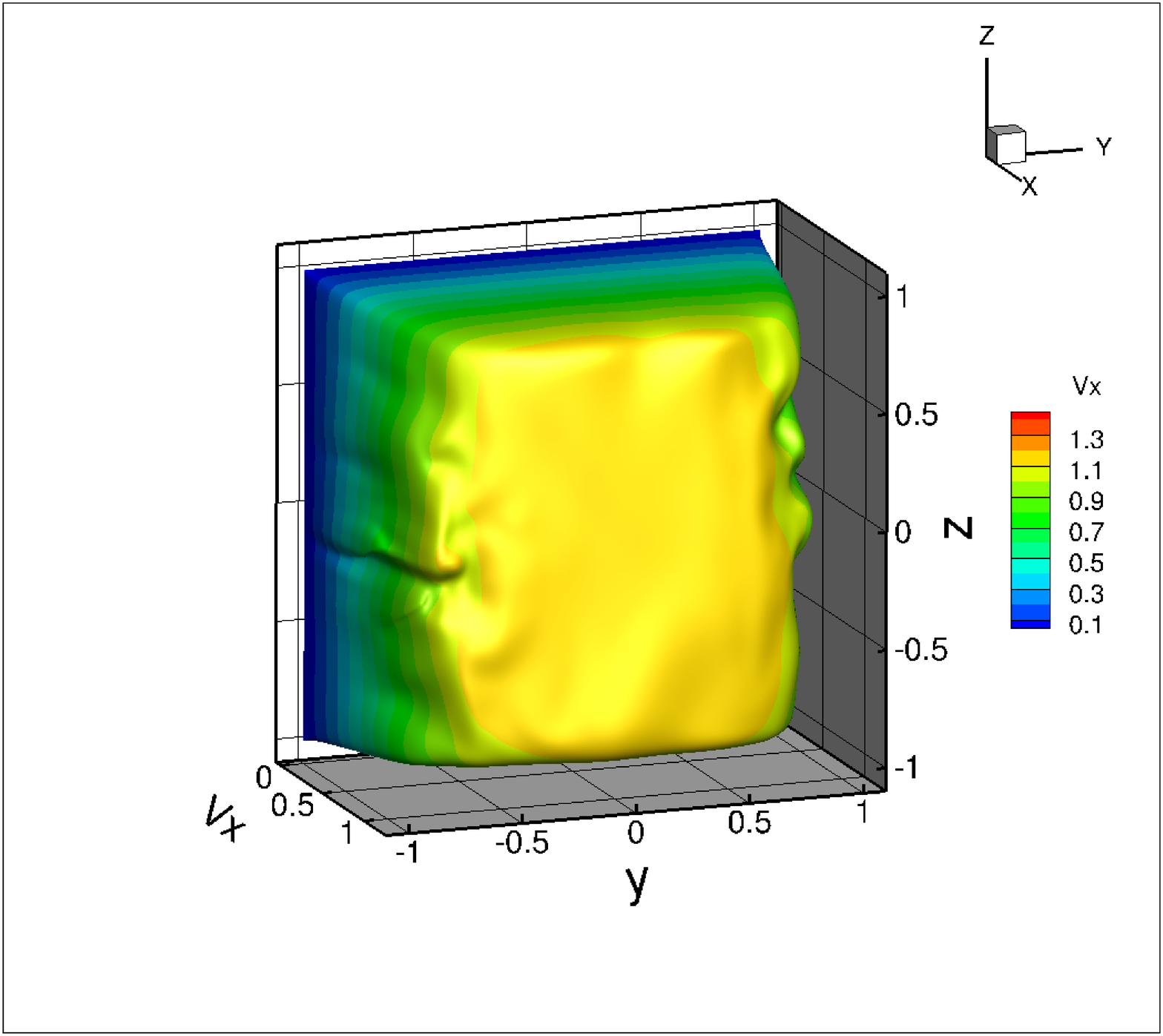}
                 \label{fig:rm0_ha28}
        }
        \subfigure[]{
                \includegraphics[width=0.42\textwidth,trim=55mm 20mm 50mm 62mm,clip]{rm100_ha28_gimp.eps}
                 \label{fig:rm100_ha28}
        }
        \subfigure{
                \includegraphics[width=0.12\textwidth,trim=320mm 90mm 5mm 112mm,clip]{rm100_ha28_gimp.eps}
        }
\caption{Instantaneous axial velocity profiles at $x=l_{x}/2$. (a) Initial state at $t=0$ without the magnetic field, (b) $Ha=26$, $R_{m}=0$ at $t=122$, the flow is almost
laminarized, (c) $Ha=26$, $R_{m}=100$ at $t=122$, the Shercliff layers continue to be turbulent, (e) $Ha=28$, $R_{m}=0$ at $t=46$, (f)  $Ha=28$, $R_{m}=100$ at $t=46$.}
\label{fig:usnap_relam}
\end{figure}
where the velocity fluctuation $\bm{v'}=[v'_{x},v'_{y},v'_{z}]$ is defined as
\begin{equation}
\bm{v'}=\bm{v}-{l_{x}}^{-1}\int \limits_{0}^{l_{x}}\bm{v}dx.
\end{equation}
Furthermore, the flow is also observed to laminarize for all cases of $Ha>26$, for example at $Ha=28$ shown in Fig.~\ref{fig:ek_relam}.

However, when $R_{m}=100$ the flow remains turbulent at $Ha=26$, shown by the settling of $\langle E_{k}\rangle$ in contrast to the QS case. This can also be
observed from the instantaneous axial velocity profiles shown in Fig.~\ref{fig:rm0_ha26} and Fig.~\ref{fig:rm100_ha26} where the QS case shows almost complete laminarization at $t=122$ 
while the $R_{m}=100$ case shows 
turbulent Shercliff layers. Neverthless, with a slightly 
stronger magnetic field ($Ha=28$), at $R_{m}=100$, turbulence is completely suppressed as in the QS case but with a slower rate of decay. This is further evident from 
Fig.~\ref{fig:rm0_ha28} and Fig.~\ref{fig:rm100_ha28}, with the $R_{m}=100$ case showing much higher intensity of turbulence in the Shercliff layers at $t=46$ as compared to the QS case. From these observations, it seems very likely that 
in the low $\Rey$ regime, a higher $R_{m}$ tends to sustain turbulence and hence delays the laminarization threshold to a higher value of Hartmann number. This behaviour can be attributed to
 the independent dynamics of the magnetic field that reduce dissipation and hence delays the energy decay. 

\subsection{Turbulence at lower Hartmann number}
\begin{figure}[!h]
      \subfigure[]{
      \includegraphics[width=0.5\textwidth]{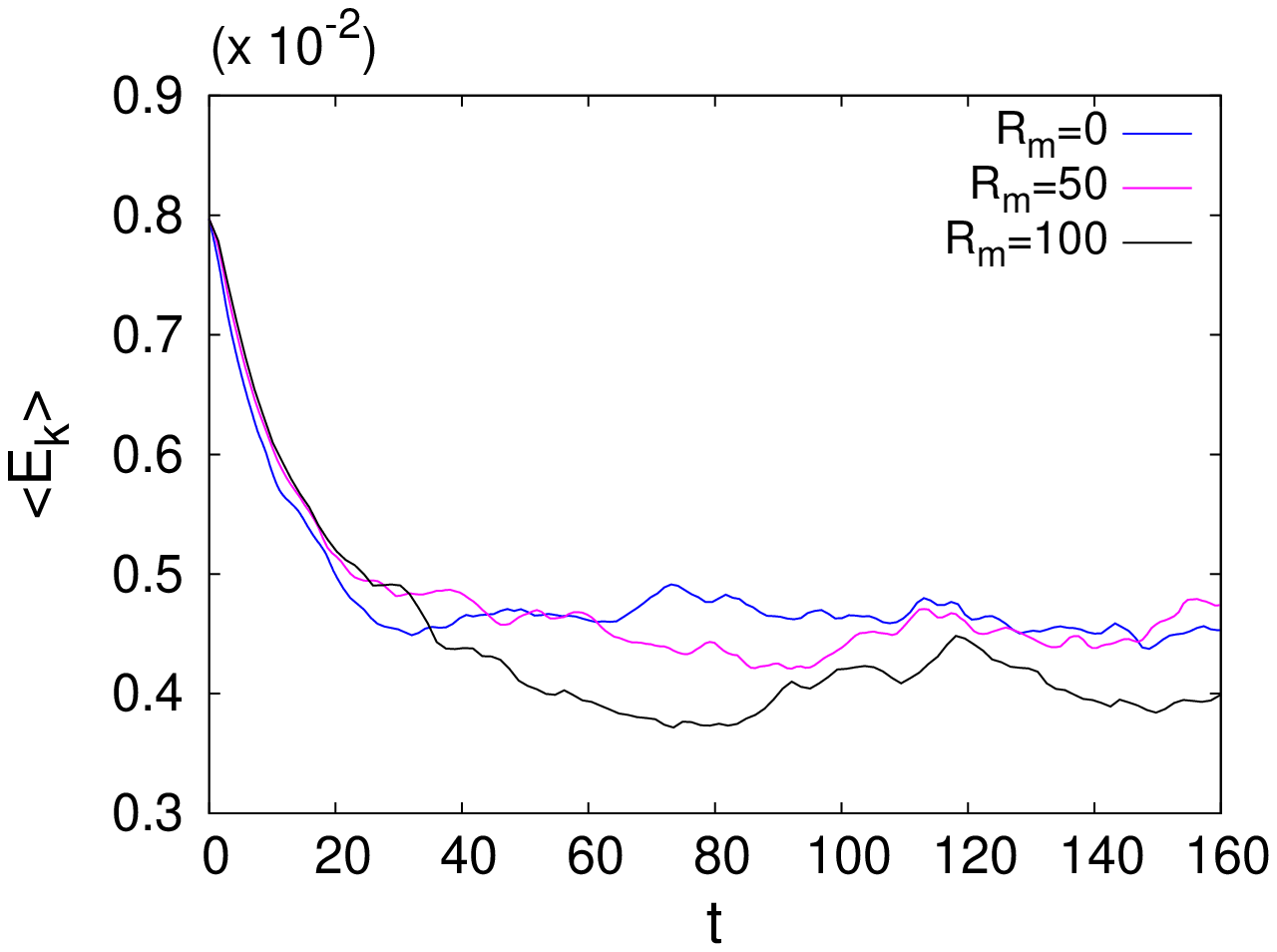}
      \label{fig:energy}
      }
      \subfigure[]{
      \includegraphics[width=0.5\textwidth]{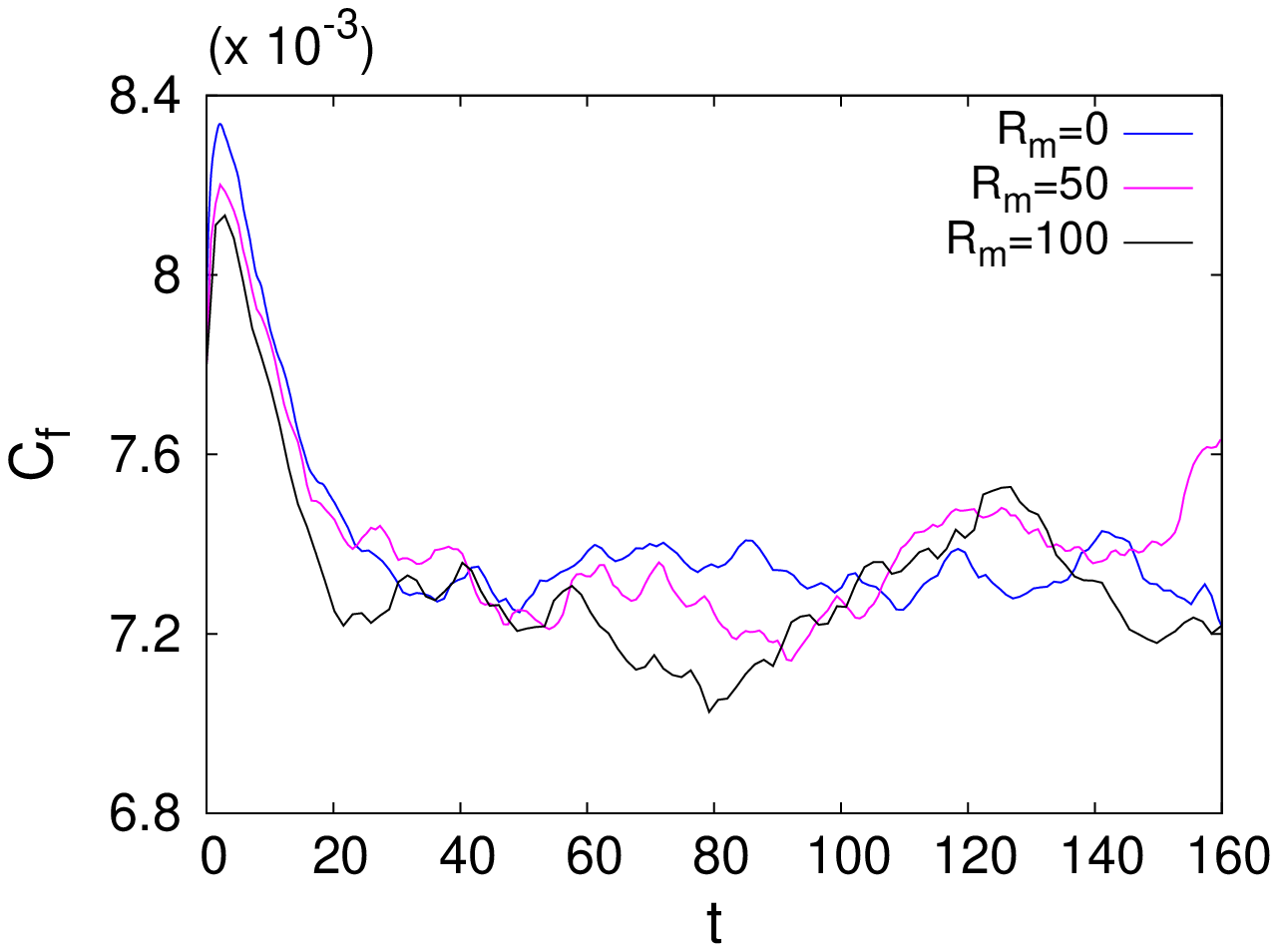}
      \label{fig:tauw}
      }
\caption{Time evolution of (a) turbulent kinetic energy $\langle E_{k}\rangle$ and (b) skin friction coefficient $C_{f}$.}
\label{fig:energytauw}
\end{figure}

In this subsection, we discuss a few features of the evolution of the turbulent flow at a relatively low Stuart number, corresponding to a Hartmann number $Ha=15$. This is performed at $R_{m}=50$ and $R_{m}=100$
along with the quasistatic case. In Fig.~\ref{fig:energy}, the decay of turbulent kinetic energy with time is shown until $t=140$. It is observed that the initial phase until around $t=30$ shows
a lower decay rate with higher $R_{m}$. However, the flow apparently reaches a statistically steady state earlier in the quasistatic case as compared to the cases with higher $R_{m}$, which show
a low frequency oscillatory behaviour after the initial steep transients. This is in line with the findings of Knaepen et. al. \cite{
Knaepen:2004} through DNS at $1\le R_{m}\le20$ in a periodic box. 

In addition, the evolved state seems to settle at turbulent energy levels roughly the same, independent of $R_{m}$. The reason for this 
is not fully clear and could be attributed to a low value of Stuart number or to the low variability in the magnetic field allowed due to the uniformity in the intial magnetic field.

It is known from quasistatic MHD that the Hartmann flow has two opposing effects namely the Hartmann flattening and turbulence suppression effects that determine the evolution of skin friction coefficient under the application of
a magnetic field. This can be observed from 
\begin{figure}[!h]
       % \begin{center}
        \subfigure[]{
                \includegraphics[width=0.54\textwidth,trim=5mm 0mm 5mm 27mm,clip]{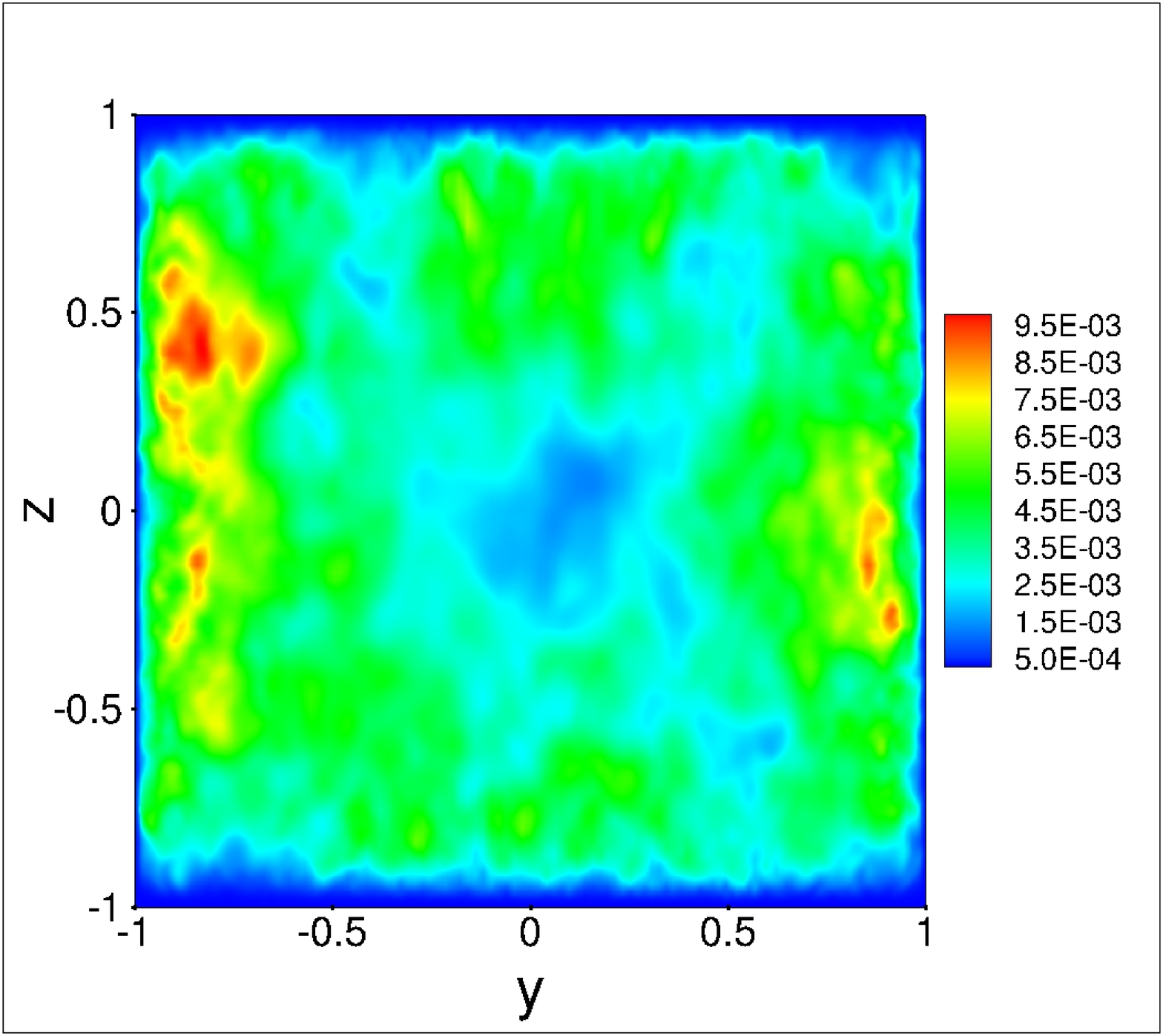}
                 \label{fig:rm0_ha0}
        }
       % \end{center} 
        \subfigure[]{
                \includegraphics[width=0.45\textwidth,trim=5mm 4mm 73mm 25mm,clip]{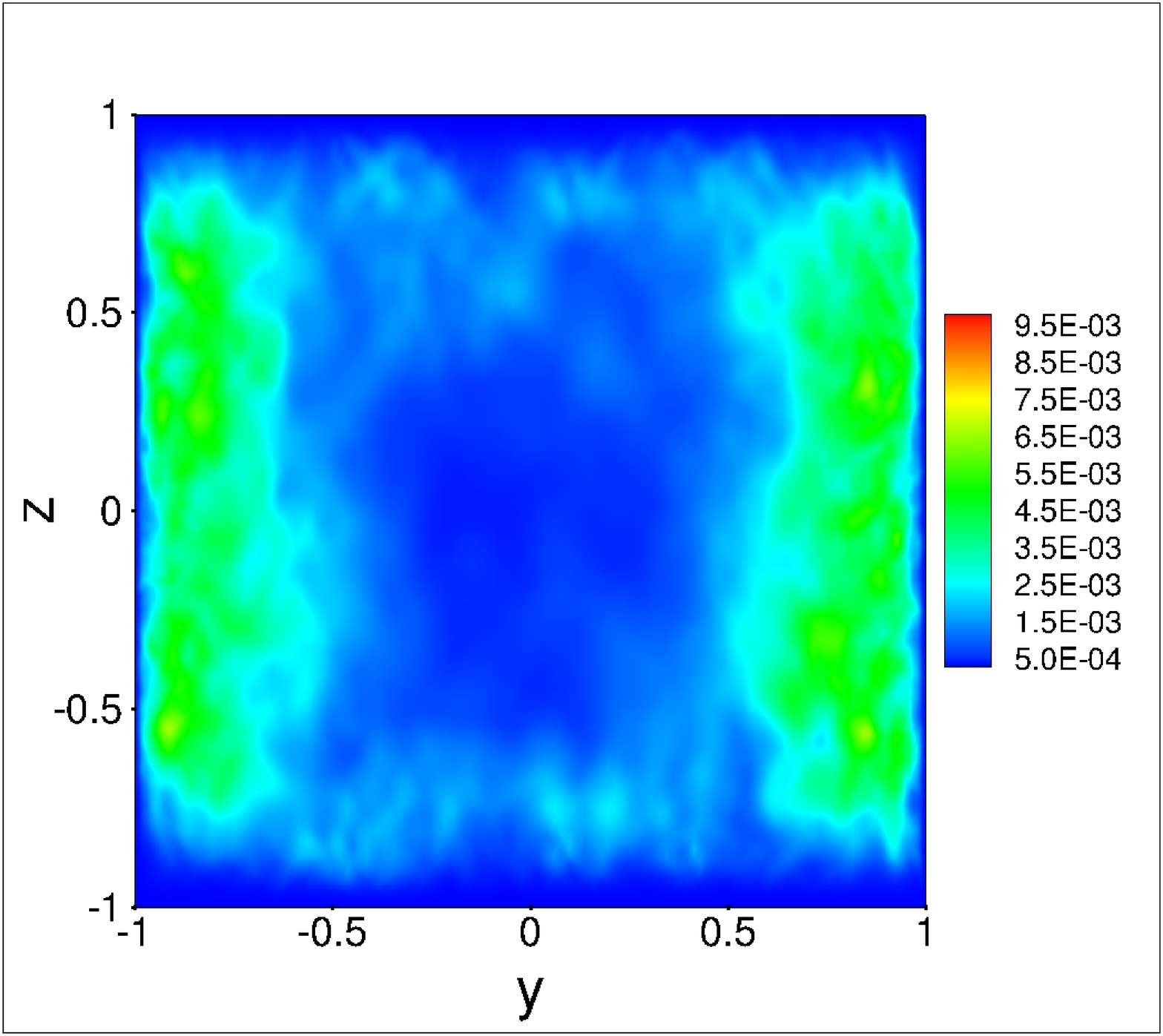}
                 \label{fig:rm0_ha15}
        }
        \subfigure[]{
                \includegraphics[width=0.45\textwidth,trim=9mm 4mm 66mm 25mm,clip]{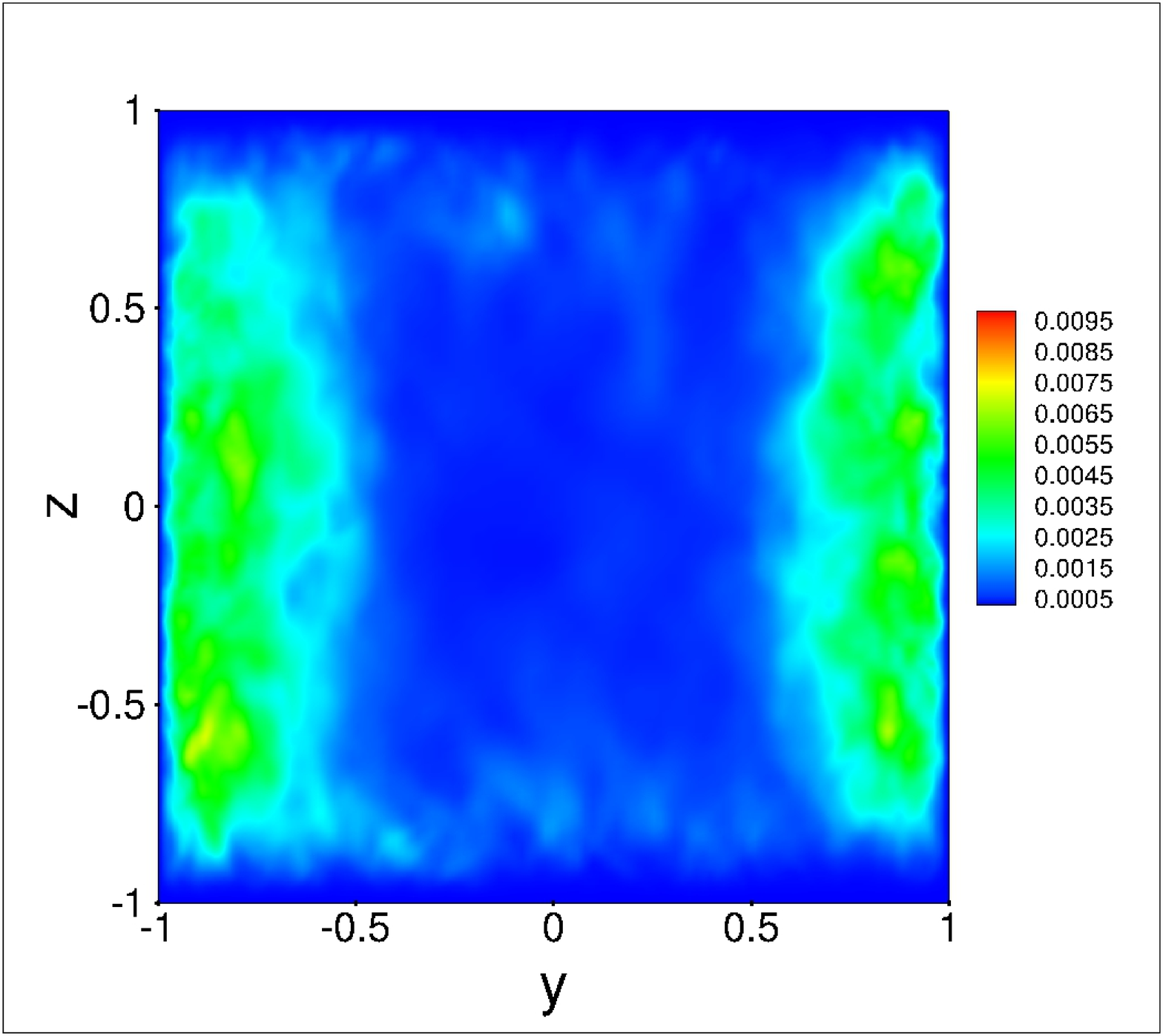}
                 \label{fig:rm50_ha15}
        }
        \qquad \quad
        \subfigure[]{
                \includegraphics[width=0.45\textwidth,trim=5mm 4mm 73mm 25mm,clip]{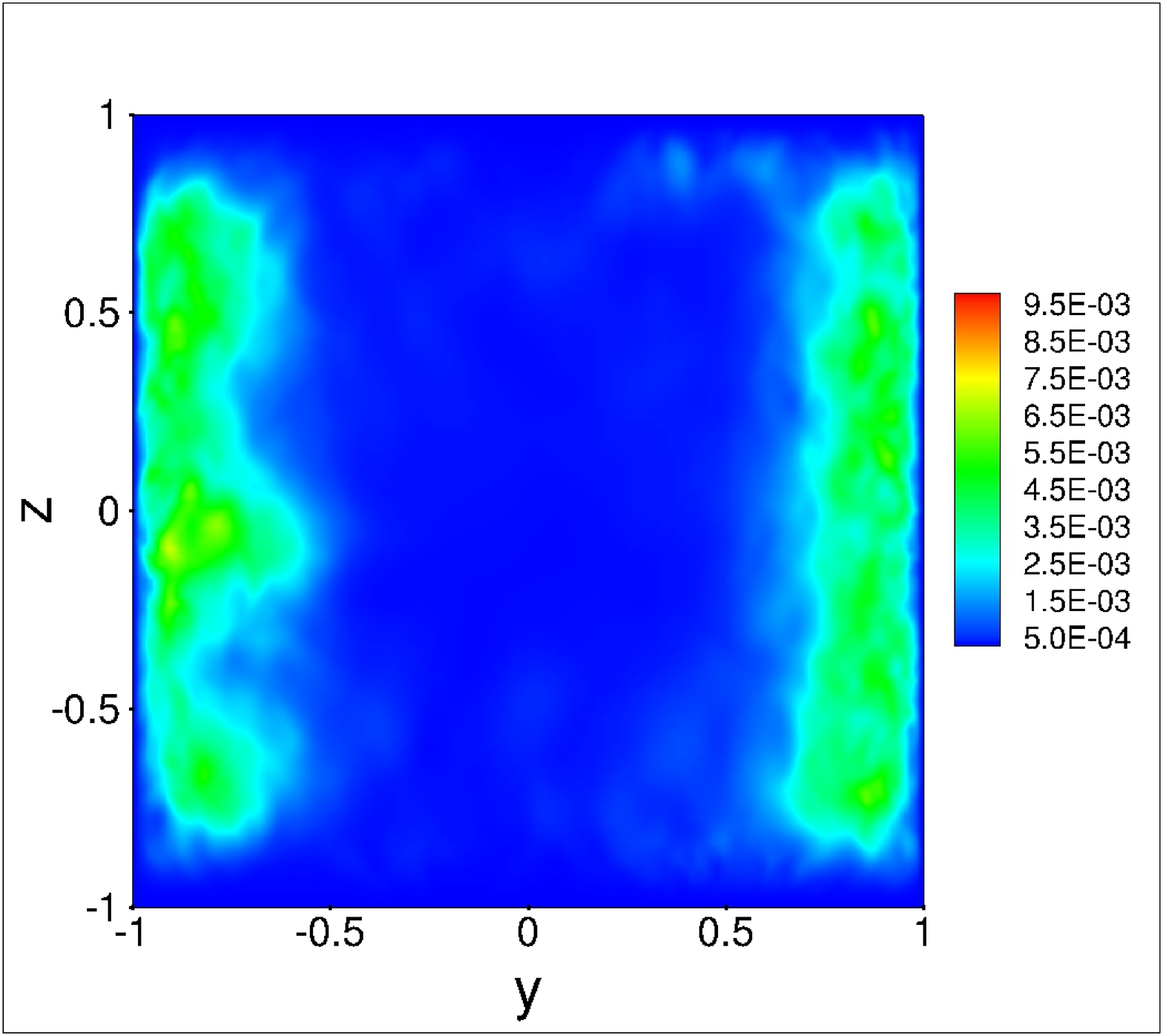}
                 \label{fig:rm100_ha15}
        }
\caption{Suppression of the Reynolds stress component $\overline{{v_{z}'}^{2}}$ due to the applied magnetic field at $t=80$. (a) Common to both cases at $t=0$; (b) quasistatic case, $R_{m}=0$, 
(c) $R_{m}=50$ and (d) $R_{m}=100$.}
\label{fig:t33_ha15}
\end{figure}
the corresponding behaviour of skin friction coefficient defined as 
\begin{equation}
\label{eq:tkenergy} C_{f}=\frac{1}{8\Rey} \int\limits_{\varGamma}-\frac {\partial \bar{v}_{x}}{\partial n} dl,
\end{equation} 
($\bar{v}_{x}$ being the 
mean axial velocity and $n$ the wall normal direction) which is shown in Fig.~\ref{fig:tauw}. An initial increase in $C_{f}$ occurs until around $t=2.3$ due to the dominance of the Hartmann flattening followed by a decease in $C_{f}$ when suppression of 
turbulence becomes important. Such a qualitative behaviour is unaffected by the magnetic Reynolds number, although a slight increase in the magnitute of the peak is observed. Finally, the
oscillatory behaviour at higher $R_{m}$ is also seen with $C_{f}$ as in the case of turbulent kinetic energy. 

The effect of $R_{m}$ on the supression of turbulence by the magnetic field can be observed from the instantaneous cross-sectional distribution of Reynolds stress tensor 
component $\overline{{v_{z}'}^{2}}$ (the overbar indicates streamwise averaging), shown at $t=80$ in Fig.~\ref{fig:t33_ha15}. Our first studies indicate clearly that with increasing $R_{m}$, the stress component is increasingly supressed 
in the core region close to the Hartmann layer.
%However, in order to make clear conclusions it will be necessary to obtain statistically steady turbulence at several values of interaction parameters. 

\section{Conclusions}
A semi-implicit numerical procedure for solving the magnetic induction equation coupling the integral boundary element method with local finite differences in the interior has been 
presented for finite $R_{m}$ MHD duct flows with periodic conditions in the streamwise direction. Specific features arising due to this configuration like corner singularities and the 
divergence of zero mode of the magnetic field ($k=0$) required special attention. Detailed verification of the numerical procedure was made in the limiting case of low $R_{m}$. The method
 is a useful tool to perform direct numerical simulations of MHD turbulence with uniform magnetic fields as well as localized magnetic fields that have been of recent interest. The 
pseudo-vacuum magnetic boundary conditions, that are typically used in commercial codes, are found to accurately describe the current density field $\bm{j}$ in the limit of low $R_{m}$, 
although the magnetic fields show sigificant differences with those computed with the integral boundary conditions. Finally, DNS runs of turbulence evolution of Hartmann duct 
flow at low $\Rey$ and moderate $R_{m}$
already point to some interesting and significant differences with the quasistatic MHD, like delaying relaminarization at relatively low Stuart numbers. We can thus expect that at
higher Stuart numbers and correspondingly higher $\Rey$, more significant differences will be observed.     
 
The present work is considered to be a starting point for the study of finite $R_{m}$ MHD turbulence in duct flows. Plans for future research include extension of this procedure to incorporate 
non-periodic inlet-outlet boundary conditions, which would enable shorter streamwise duct lengths required for turbulence studies. The present model assumes the absence of a net mean 
streamwise current which is typical of most applications. However, incorporating this could be of some theoretical interest to enable computation of flows with imposed streamwise currents. 
For liquid metals typically $Pr_{m}\sim10^{-6}$ which would lead to very high $\Rey$ corresponding to $R_{m}\sim1$. Modelling turbulence at such a high $\Rey$ through direct numerical 
simulations is computationally not feasible due to the high grid resolutions required to resolve the small velocity scales. Our present focus has been on relatively high magnetic Prandtl 
numbers $Pr_{m}\geq10^{-3}$. Modeling flows with realistic magnetic Prandtl numbers ($Pr_{m}\sim10^{-6}$) would be feasible with the use of large eddy simulations (LES) that requires 
subgrid-scale turbulence modeling coupled with the boundary integral procedure which we plan to pursue in the near future.

%%%%%%%%%%%%%%%%%%%%%%%%%%%%%%%%%%%%%%%%%%%%%%%
% Acknowledgements %
\section*{Acknowledgments}
We acknowledge financial support from the Deutsche Forschungsgemeinschaft within Research Training Group 1567 and the Helmholtz Alliance ``Liquid Metal Technologies''. 
VB gratefully acknowledges fruitful discussions with Andre Giesecke from Helmholtz-Zentrum Dresden-Rossendorf (HZDR), Germany. Computer resources were provided by
the computing center of TU Ilmenau.
%%%%%%%%%%%%%%%%%%%%%%%%%%%%%%%%%%%%%%%%%%%%%%%

\section{Appendix}

\subsection{Uniqueness of the current density field $\bm{j}$ when $j_{n}=0$ in the QST formulation ($R_{m}\ll1$)}

The current density field is observed to be independent of the exact form of the magnetic boundary conditions which satisfy $j_{n}=0$, to which we provide a simple proof. Considering two numerical realizations of computing the secondary magnetic field $\bm{b}$ from the QST formulation with a given velocity field $\bm{v}$ , we denote the solutions obtained as $\bm{b}_{1}$ and $\bm{b}_{2}$. As an example, one of the cases can correspond to the integral boundary conditions shown in this paper and the other case can correspond to the pseudo-vacuum boundary conditions. Both boundary conditions ensure that the wall normal current vanishes, $j_{n}=0$. 
The difference between the two solutions is denoted by $\bm{d}_{b} = \bm{b}_{2} - \bm{b}_{1}$. It follows from equation \eqref{quasistationary} that 
% laplacian of e=0
\begin{equation}
\nabla^2 \bm{d}_{b}= 0 \hskip4mm \textrm{or} \hskip4mm \nabla\left(\nabla\cdot \bm{d}_{b} \right) - \nabla \times \left(\nabla \times \bm{d}_{b}\right) = 0 \textrm{.}
\end{equation}
Since both the solutions $\bm{b}_{1}$ and $\bm{b}_{2}$ are solenoidal, $\nabla \times \left(\nabla \times \bm{d}_{b}\right) = 0$ and we can introduce a scalar potential $\phi$ as $\nabla\times\bm{d}_{b}=-\nabla \phi$. Taking the divergence, we obtain 
% laplacian of phi = 0
\begin{equation}
\label{poissonphi} \nabla^2 \phi= 0 \textrm{.} 
\end{equation}
Moreover, $ \nabla \phi =  -\nabla \times \bm{d}_{b} = -\nabla \times \left(\bm{b}_{2} - \bm{b}_{1} \right)  = \bm{j}_{1} - \bm{j}_{2} $
Therefore it follows that $\frac{\partial \phi}{\partial n} = j_{n1} - j_{n2} = 0$ on the boundary.
Equation \eqref{poissonphi} with the Neumann condition gives $\phi=constant$ and hence $\bm{j}_{1} = \bm{j}_{2}$. This explains why the solution for $\bm{j}$ obtained from the pseudo-vacuum boundary conditions is in agreement with that obtained using the rigorous boundary integral procedure.

\subsection{Effective wavenumbers for Fast Fourier transformation}

The system considered here is periodic in the $x$-direction and hence FFT is applied in that direction. This enables efficient parallelization through the solution of Poisson (elliptic equations, in general) equations for velocity ($\bm{v}$), pressure ($p$) and the mean streamwise magnetic field ($\bar{b}_x$) using the Fishpack 2D solver. In addition, parallelization of the coupled BEM procedure with non-local boundary conditions becomes much easier without inter-processor communication. However, the streamwise derivatives computed in the Fourier space are not equivalent to the derivatives approximated using finite differences. Equivalence can be attained by the use of effective wavenumbers $\alpha_{k1}$ and $\alpha_{k2}$ that correspond to the first and second $x$-derivatives respectively as

% alpha effective
\begin{equation}
\alpha_{k1} = \frac{\sin{\left( \alpha\delta x\right) }}{\delta x} \textrm{,} \hskip3mm  \hskip3mm \alpha_{k2} = \frac{\sin{\left( \alpha\frac{1}{2}\delta x\right)  }}{\frac{1}{2}\delta x} \textrm{.}
\end{equation}
\cite {Ferziger-book}. These relations are obtained by substituting the function $e^{i \alpha_{k} x}$ into the central finite-difference stencils for the first and second derivatives, respectively. It must be mentioned that in our procedure, $\alpha_{k1}$ is applied only in reconstructing the streamwise component $\hat{b}_{x}$ from $\hat{b}_{y}$ and $\hat{b}_{z}$ from the divergence-free condition using 
\begin{equation}
\hat{b}_{x} = \frac{-1}{i\alpha_{k1}}\left(\frac{\partial \hat{b}_{y}}{\partial y} + \frac{\partial \hat{b}_{z}}{\partial z}\right),\:k\neq0 
\end{equation}
and $\alpha_{k2}$ is used for the rest of the procedure. 

\subsection{Effect of boundary conditions on the generation of $\nabla\cdot\bm{b}$ in the semi-implicit evolution of $\bm{b}$}

Although Faraday's law which is given by 
\begin{equation}
\frac{\partial\bm{b}}{\partial t}= -\nabla\times \bm{e}
\end{equation}
dictates that the divergence of the evolving magnetic field must vanish analytically, the corresponding semi-implicit finite-difference procedure leads to divergence issues due to the presence of boundaries. Considering a simple 2D uniform rectangular grid of size $h$, we now show how the boundary conditions lead to the generation of $\nabla\cdot\bm{b}$. Choosing $i$ and $j$ as the grid indices along the $y$ and $z$ directions respectively, the second-order discrete 2D Laplace operator for a general scalar $\phi$ can be written as
\begin{equation}
\nabla^2\phi = \frac{\phi_{i+1,j} -2\phi_{i,j} + \phi_{i-1,j} }{h^2} + \frac{\phi_{i,j+1} -2\phi_{i,j} + \phi_{i,j-1} }{h^2}
\end{equation}

The discrete 2D version of equation \eqref{bpoisson} can be written as

\begin{equation}
\begin{split}
-f {b_{y}}^{n+1}_{i+1,j} + \frac{{b_{y}}^{n+1}_{i+2,j} -2{b_{y}}^{n+1}_{i+1,j} + {b_{y}}^{n+1}_{i,j} }{h^2} +  \frac{{b_{y}}^{n+1}_{i+1,j+1} -2{b_{y}}^{n+1}_{i+1,j} + {b_{y}}^{n+1}_{i+1,j-1} }{h^2} \\ = -f {q_{y}}_{i+1,j}
\end{split}
\end{equation}
\begin{equation}
\begin{split}
-f {b_{y}}^{n+1}_{i-1,j} + \frac{{b_{y}}^{n+1}_{i,j} -2{b_{y}}^{n+1}_{i-1,j} + {b_{y}}^{n+1}_{i-2,j} }{h^2} +  \frac{{b_{y}}^{n+1}_{i-1,j+1} -2{b_{y}}^{n+1}_{i-1,j} + {b_{y}}^{n+1}_{i-1,j-1} }{h^2} \\ = -f {q_{y}}_{i-1,j}
\end{split}
\end{equation}
\begin{equation}
\begin{split}
\label{topwall} -f {b_{z}}^{n+1}_{i,j+1} + \frac{{b_{z}}^{n+1}_{i+1,j+1} -2{b_{z}}^{n+1}_{i,j+1} + {b_{z}}^{n+1}_{i-1,j+1} }{h^2} +  \frac{{b_{z}}^{n+1}_{i,j+2} -2{b_{z}}^{n+1}_{i,j+1} + {b_{z}}^{n+1}_{i,j} }{h^2} \\ = -f {q_{z}}_{i,j+1}
\end{split}
\end{equation}
\begin{equation}
\begin{split}
-f {b_{z}}^{n+1}_{i,j-1} + \frac{{b_{z}}^{n+1}_{i+1,j-1} -2{b_{z}}^{n+1}_{i,j-1} + {b_{z}}^{n+1}_{i-1,j-1} }{h^2} +  \frac{{b_{z}}^{n+1}_{i,j} -2{b_{z}}^{n+1}_{i,j-1} + {b_{z}}^{n+1}_{i,j-2} }{h^2} \\ = -f {q_{z}}_{i,j-1}
\end{split}
\end{equation}
Performing (61)-(62)+(63)-(64) and dividing by $2h$ gives
\begin{equation}
\label{dpoisson} -f d^{n+1}_{i,j} + \nabla^2 d^{n+1}_{i,j} = -f {d_{q}}_{i,j} = 0 \textrm{,}
\end{equation}
where $d$ and $d_{q}$ represent the divergence of $\bm{b}^{n+1}$ and $\bm{q}$ respectively. The term $\bm{q}$ being a resultant of linear operations on the advective and field stretching terms $(\bm{v}^n\cdot \nabla)\bm{b}^n_{t}$ and $(\bm{b}^n_{t}\cdot \nabla)\bm{v}^n$ can be shown to be divergence free. The boundary condition for equation \eqref{dpoisson} is the Dirichlet condition $d^{n+1}=0$ on $y,z=\pm1$ which follows directly from the application of \eqref{bneqn} in the 2D version. However, equation \eqref{dpoisson} is applicable only when the point $(i,j)$ under consideration is not adjacent to the boundary. In the case when $(i,j)$ is adjacent to the boundary (say when $j=j_{max}-1$ which corresponds to a grid point just below the top wall), equation \eqref{topwall} must be replaced by a discrete operator $\pounds$ consisting of equations \eqref{bie}, \eqref{bteqn} and \eqref{bneqn} as
\begin{equation}
{b_{z}}^{n+1}_{i,j+1}  = \pounds\left\lbrace {b_{z}}^{n+1}_{i,j},\bar{b}_{\tau}\right\rbrace \textrm{,} 
\end{equation}
where $\bar{b}_{\tau}$ is the vector containing all the tangential components of magnetic field on the rectangular boundary $\varGamma$. This clearly shows that equation \eqref{dpoisson} does not hold for the interior grid points adjacent to $\varGamma$ which forms the numerical source of divergence and diffuses into the entire interior according to equation \eqref{dpoisson}. Due to this reason, reconstruction of $b_{x}$ for non-zero modes and the use of vector potential is essential when the diffusive term in the induction equation is treated implicitly. This however is not the case in a fully explicit scheme (typically used when the diffusive time scales are much higher, $R_{m}\gg1$) as the necessity to solve the Poisson equation does not arise at all \cite{Iskakov:2004}.

%%%%%%%%%%%%%%%%%%%%%%%%%%%%%%%%%%%%%%%%%%%%%%%
% Bibliography %
%\section*{References}
\bibliographystyle{elsarticle-num}
\bibliography{refs.bib}
\biboptions{sort&compress}
%\nocite{*}
%%%%%%%%%%%%%%%%%%%%%%%%%%%%%%%%%%%%%%%%%%%%%%%

\end{document}